\PassOptionsToPackage{table}{xcolor}
\documentclass[sigconf]{acmart}

\AtBeginDocument{%
  }

\copyrightyear{2026}
\acmYear{2026}
\setcopyright{cc}
\setcctype{by}
\acmConference[CHI '26]{Proceedings of the 2026 CHI Conference on Human Factors in Computing Systems}{April 13--17, 2026}{Barcelona, Spain}
\acmBooktitle{Proceedings of the 2026 CHI Conference on Human Factors in Computing Systems (CHI '26), April 13--17, 2026, Barcelona, Spain}
\acmDOI{10.1145/3772318.3790266}
\acmISBN{979-8-4007-2278-3/2026/04}

\usepackage{fancyvrb}
\usepackage{xcolor}
\usepackage{kotex}
\usepackage{multirow}
\usepackage{array}
\usepackage{colortbl}

\begin{document}

\title{AI Twin: Enhancing ESL Speaking Practice through AI Self-Clones of a Better Me}

\author{Minju Park}
\authornote{Equal contribution.}
\email{minju.park@ubc.ca}
\affiliation{%
  \institution{University of British Columbia}
  \city{Vancouver}
  \state{BC}
  \country{Canada}
}
\author{Seunghyun Lee}
\authornotemark[1]
\email{seunghyun.lee@socra.ai}
\affiliation{%
  \institution{Socra AI}
  \city{Seoul}
  \country{South Korea}
}
\author{Juhwan Ma}
\email{juhwan.ma@socra.ai}
\affiliation{%
  \institution{Socra AI}
  \city{Seoul}
  \country{South Korea}
}
\author{Dongwook Yoon}
\email{yoon@cs.ubc.ca}
\affiliation{%
  \institution{University of British Columbia}
  \city{Vancouver}
  \state{BC}
  \country{Canada}
}

\renewcommand{\shortauthors}{Park et al.}

\begin{abstract}
Advances in AI have enabled ESL learners to practice speaking through conversational systems. However, most tools rely on explicit correction, which can interrupt the conversation and undermine confidence. Grounded in second language acquisition and motivational psychology, we present \emph{AI Twin}, a system that rephrases learner utterances into more fluent English and delivers them in the learner's voice. Embodying a more confident and proficient version of the learner, AI Twin reinforces motivation through alignment with their aspirational \emph{Ideal L2 Self}. Also, its use of implicit feedback through rephrasing preserves conversational flow and fosters an emotionally supportive environment. In a within-subject study with 20 adult ESL learners, we compared AI Twin with explicit correction and a non-personalized rephrasing agent. Results show that AI Twin elicited higher emotional engagement, with participants describing the experience as more motivating. These findings highlight the potential of self-representative AI for personalized, psychologically grounded support in ESL learning.
\end{abstract}




\begin{teaserfigure}
  \includegraphics[width=\textwidth]{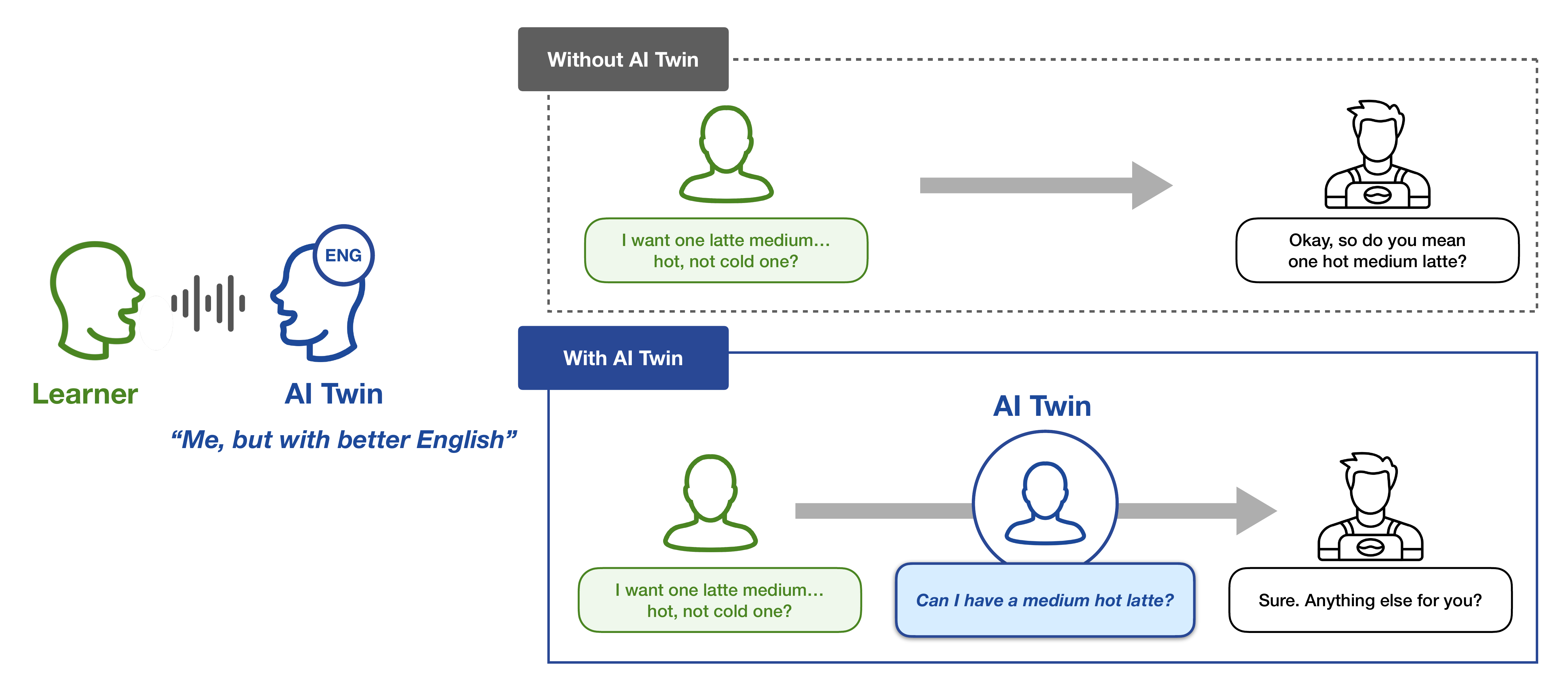}
  \caption{\emph{AI Twin} represents a self-clone---a ``better me'' that speaks more fluent English---supporting ESL learners by rephrasing their speech in more natural English using their own voice. In conversation practice with an AI interlocutor, AI Twin transforms the learner's utterance into clearer English, enabling smoother and more successful practice interactions.}
  \Description{The figure has two parts. On the left, it visualizes the concept of AI Twin as a self-clone of the learner that speaks more fluent English. On the right, it shows example conversation scenarios. The top row illustrates practice without AI Twin, where the learner's utterance remains unmodified. The bottom row illustrates practice with AI Twin, where the learner's utterance is rephrased into clearer English in their own voice, supporting smoother interaction with the AI interlocutor.}
  \label{fig:teaser}
\end{teaserfigure}

\maketitle

\section{Introduction}
Artificial intelligence (AI) has long been explored as a means to support English as a Second Language (ESL) learning, offering opportunities for conversational practice beyond the classroom~\cite{sari2023role}.
Earlier systems, from text-based chatbots~\cite{fryer2006bots} to speech-enabled tools~\cite{ehsani1998speech, seneff2004spoken}, attempted to provide such practice but were constrained by scripted responses and narrow task coverage~\cite{raux2004using}.
Recent advances in generative artificial intelligence (GenAI)---particularly large language models (LLMs) and neural speech synthesis---now overcome these limitations, enabling open-ended spoken dialogue that adapts to learners' input in real-time and more closely approximates authentic communication~\cite{law2024application}.
These advances enable learners to interact with AI chatbots in more engaging situational dialogues and role-play scenarios~\cite{huang2022chatbots}, making conversational practice more realistic and immersive, and opening new possibilities for language learning.

Most AI-assisted ESL speaking tools deliver corrective feedback in addition to simulating conversation~\cite{zou2023investigation, ericsson2023english}.
However, emphasizing corrective feedback, particularly in explicit forms, can create drawbacks.
For example, providing too many corrections may discourage learners~\cite{vengadasamy2002responding}, and negatively affect emotions and motivation~\cite{yu2021works}.
These emotional effects can discourage risk-taking, limit opportunities for practice, and ultimately slow language acquisition~\cite{dewaele2007predicting,yang2024reducing}.
Although conversational AI systems are often valued for lowering anxiety compared to human interlocutors~\cite{zhang2024artificial, tai2023impact, wang2024impact}, their current feedback mechanisms may reintroduce pressure and inadvertently counteract the very benefits they are intended to provide.
This highlights the need for feedback approaches that not only promote accuracy but also provide emotional and motivational support.
Krashen's \emph{Affective Filter Hypothesis} underscores the importance of affect and motivation in language acquisition~\cite{krashen1982principles}. 
Engagement shows similar importance in online learning, where higher engagement is consistently associated with better learning outcomes~\cite{rajabalee2020study, kuzminykh2021relationship}. 
Since existing feedback mechanisms offer limited support for engagement, this gap represents an opportunity to enrich AI-assisted speaking environments with strategies that create a more motivating and emotionally supportive interaction environment.

Given this need for feedback approaches that balance accuracy with emotional and motivational support, a frequently discussed alternative in second language acquisition (SLA) literature is the use of recasts. Recasts provide implicit correction by reformulating a learner's utterance---for example, if a learner says ``She go to work'', the teacher might respond ``Yes, she goes to work''~\cite{long1998role}. Recasts preserve conversational flow and can reduce anxiety compared to explicit correction, but their subtlety often makes them ambiguous, leaving learners unsure whether a correction has been given~\cite{loewen2006recasts, lyster1998negotiation, panova2002patterns}.

To address the challenges of providing corrective feedback that both supports motivation and sustains conversational immersion, we draw on motivational psychology and the concept of the \emph{Ideal L2 Self}. The Ideal L2 Self describes a learner's envisioned, idealized future self as a proficient second language user, which has been shown to serve as a powerful motivator for language learning~\cite{dornyei2005psychology}.
By making learners aware of the gap between their current and ideal selves~\cite{higgins1987self}, it encourages them to engage in more goal-directed behaviors~\cite{markus1989possible}.
Building on this insight, we envision a conversational system designed not only to provide corrective support but also to enhance learners' motivation by enabling learners to engage with a chatbot through their idealized L2 self. By situating reformulations within a learner's own self-clone, our design preserves the advantages of implicit feedback while making its corrective function more salient and motivationally aligned.

To realize this vision, we introduce \emph{AI Twin} (Figure~\ref{fig:teaser}), a personalized self-clone---a ``better me'' that speaks more fluent English---designed to enhance learner motivation and engagement through supportive, in-conversation language learning.
In this setting, the learner converses with an AI interlocutor, while AI Twin operates in between through in-conversation rephrasing, rendering the learner's utterances in a more fluent and confident form spoken in the learner's cloned voice before passing them on. In doing so, AI Twin embodies the learner's aspirational self-image by speaking as a more proficient version of them, aiming to support learner engagement. By rephrasing rather than explicitly correcting, it maintains conversational immersion while reducing performance pressure, creating a more supportive and motivating interaction environment.
We view this focus on engagement as pedagogically meaningful, as SLA and online learning research consistently link engagement to more effective language development. Accordingly, our study addresses two research questions:

\begin{itemize}
    \item \textbf{RQ1.} How does AI-mediated in-conversation rephrasing of learner utterances, compared to explicit correction, influence ESL learners' emotional, cognitive, and behavioral engagement?
    \item \textbf{RQ2.} How does aligning rephrasing with the learner's aspirational self-image through AI Twin further shape these outcomes?
\end{itemize}

To investigate these questions, we conducted a within-subject study with 20 adult ESL learners, comparing the proposed AI Twin to two baselines: Explicit Feedback, where the system directly corrected learner utterances, and AI Proxy, where rephrased utterances were delivered in a non-personalized voice.
Our mixed-method design combined engagement surveys and semi-structured interviews to examine emotional, cognitive, and behavioral engagement across conditions. 
This approach allowed us to examine the effects of AI Twin and to disentangle whether its impact stemmed from feedback style or from personalization, drawing on both quantitative patterns and qualitative reflections on learner experience. Results showed that rephrasing fostered significantly higher emotional engagement than providing explicit feedback, and AI Twin further strengthened this effect by fostering motivation through alignment with the learner's Ideal L2 self.

In summary, our work makes three contributions.
\begin{enumerate}
    \item  We present \emph{AI Twin}, an AI self-clone that embodies the learner's Ideal L2 Self by delivering implicit, in-conversation rephrasing in the learner's own voice.
    \item We empirically evaluate AI Twin, showing in-conversation rephrasing fosters greater engagement than explicit feedback, and that personalization through self-representation further amplifies this effect.
    \item We derive design implications for affectively supportive AI learning tools, illustrating how self-representative AI can be integrated into educational technologies to promote learner engagement.
\end{enumerate}

\section{Related Work}
\subsection{AI-Supported Conversational Practice in ESL}
Role-play activities are widely used in ESL education to practice speaking skills~\cite{maarof2018effect}, yet learners often lack opportunities for authentic conversation, which requires an available interlocutor. AI-driven chatbots have been introduced to fill this gap, evolving from rigid rule-based designs to LLM-based systems that enable more dynamic and adaptive interactions~\cite{li2022using, huang2022chatbots}.
These systems can not only act as an always-available interlocutor, increasing opportunities for practice~\cite{fryer2006bots}, but also adapt their responses to learners' proficiency levels and goals.
Moreover, through voice generation, they create more natural opportunities for oral practice.

AI-based language learning systems also hold promise for creating low-anxiety environments where learners feel more comfortable taking risks in speaking. Studies indicate that learners hesitant to interact with humans often feel more at ease communicating with computers~\cite{fryer2006bots}. Automatic speech recognition (ASR) systems have similarly encouraged learners to produce more spoken output in less stressful contexts~\cite{chen2011developing}. Conversational AI systems such as ChatGPT have broadened these possibilities, with studies showing gains in oral proficiency, greater willingness to communicate, and reduced foreign language anxiety~\cite{zheng2025influence}.

Although AI holds considerable affective potential for language learning, few system designs actively leverage it. In practice, AI-supported ESL systems still concentrate on accuracy and correction, giving far less attention to learner confidence, motivation, and willingness to experiment with language. While AI-enhanced environments can offer personalized support and alleviate emotional distress, they may also heighten anxiety or frustration if learners feel overwhelmed or unsupported~\cite{yang2025transforming}. 
Moreover, recent work suggests that although LLM-based systems achieve high accuracy, they still lack human-level cognitive capabilities~\cite{sharma2024comuniqa}. This tension underscores the importance of designs that explicitly integrate affective considerations. Our work builds on this trajectory by examining how AI-supported conversational learning systems can be designed not only to support proficiency but also to foster more engaging and supportive learning experiences.

\subsection{Corrective Feedback and Engagement in SLA}
Corrective feedback (CF) has long been recognized as a central mechanism in SLA, as it helps learners recognize errors in their language use and explore how to improve them~\cite{lyster1997corrective}. Scholars have distinguished a range of feedback types, including explicit correction, where errors are overtly identified, and implicit strategies such as recasts, in which a learner's utterance is reformulated in a more target-like way without directly flagging the mistake~\cite{sheen2011corrective}.
Explicit feedback offers clarity and reliably directs attention to errors~\cite{ellis2006implicit}, though it is often most effective when both learner and instructor are focused on form rather than engaged in ongoing conversation~\cite{oliver2003interactional}. In contrast, implicit strategies like recasts preserve conversational flow and can reduce anxiety~\cite{loewen2006recasts}, yet they may lack salience: learners sometimes fail to notice the correction~\cite{lyster1998negotiation, kim2004issues} or interpret the reformulation as mere confirmation~\cite{panova2002patterns, jang2011corrective}. This ambiguity can place additional demands on learners, making it difficult to sustain immersion in conversation.

These pedagogical insights are becoming increasingly relevant in AI-supported language learning environments. Conversational agents and chatbots now provide opportunities for tailored, automated feedback during or after interactions~\cite{li2022using}. Recent works have integrated SLA-informed CF strategies into such systems~\cite{kim2024effects, chhabriya2024ai}, and notably, recasts have been shown to remain effective in AI-mediated contexts, suggesting that implicit CF strategies may also function effectively in these environments~\cite{kim2024effects}. However, most existing AI-supported ESL systems still rely heavily on explicit error correction, which can interrupt dialogue and discourage risk-taking~\cite{zou2023investigation, ericsson2023english}. Others explore alternative architectures, such as separating conversational and feedback-providing agents in role-playing scenarios~\cite{liang2023chatback, yu2023chatlang}. Together, these systems illustrate the growing interest in embedding pedagogically grounded feedback mechanisms into AI-mediated learning.

Within this landscape, learner engagement and motivation provide an important lens for understanding the effects of CF and its design. Learner engagement is commonly defined as ``the time and effort students devote to activities that are empirically linked to desired outcomes''~\cite{kuh2009student}. In SLA, engagement has been described as the mechanism that links instructional support to actual learning outcomes~\cite{han2015exploring, shen2023learner}, and as a process that enables learners to internalize feedback~\cite{best2015listening}. Following widely used frameworks~\cite{fredricks2004school}, engagement encompasses emotional (e.g., enjoyment, anxiety), cognitive (e.g., noticing, strategic processing), and behavioral (e.g., persistence, participation) dimensions, and this tripartite framework has been widely applied in SLA and ESL contexts~\cite{ellis2010epilogue, moser2020written, zheng2018student}. However, despite its importance in learning, engagement has received comparatively limited attention in CF-focused ESL research~\cite{liu2023role}, even though the manner of feedback delivery can shape these forms of engagement~\cite{ene2021does}.

Motivation is closely intertwined with engagement and is widely recognized as an essential driver of L2 learning~\cite{ryan2000self, gardner2010motivation}.
While some scholars use the terms interchangeably~\cite{national2003engaging}, motivation can be understood as the underlying desire to learn, and engagement as the enactment of that desire in emotional, cognitive, and behavioral terms during specific learning activities. Drawing on this distinction, our study considers both motivation and engagement as central to understanding how learners interact with the AI Twin system. Specifically, we adopt the three-dimensional framework of engagement~\cite{fredricks2004school} to evaluate emotional, cognitive, and behavioral responses, while situating these within the broader motivational context of L2 learning.

\subsection{Aspirational Self-Representations}
The concept of self-representations, or \emph{possible selves}, was originally articulated in psychology by Markus and Nurius~\cite{markus1986possible} to describe the individuals one might become in the future and the self-guides that shape behavior. Higgins’ self-discrepancy theory further explains their motivational force, proposing that individuals are driven to reduce the gap between their actual self and envisioned future selves~\cite{higgins1987self}.
Applying these concepts to SLA, Dörnyei introduced the concept of the \emph{Ideal L2 Self} to describe the L2-specific dimension of one's ideal self~\cite{dornyei2005psychology}. This construct implies that when learners imagine themselves as fluent and confident L2 speakers, the Ideal L2 Self becomes a strong motivator by encouraging them to reduce the gap between who they are now and who they aspire to be.
This perspective expands traditional views of L2 motivation and resonates with other theoretical accounts that emphasize identity and long-term motivational processes~\cite{noels2003you, ushioda2001motivational}, positioning the Ideal L2 Self as a useful framework for explaining sustained language learning motivation.

Parallel developments in HCI have explored how digital representations of people---such as avatars or AI clones---can influence engagement, behavior, and self-perception. AI clones have been envisioned as digital entities that simulate a person's likeness, behavior, or decision-making style through algorithmic modeling~\cite{lee2023speculating}, often for personal assistance~\cite{personal-ai} or proxy interactions~\cite{leong2024dittos}. Advances in multimodal synthesis have increased the fidelity of these representations, making digital likenesses more realistic and salient~\cite{truby2021human}. In contrast, \emph{AI self-clones} refer to digital representations that individuals deliberately create and engage with themselves~\cite{huang2025mirror}, circumventing many ethical concerns around unauthorized mimicry~\cite{mcilroy2022mimetic} and enabling opportunities for self-reflection. Encounters with such digital selves have been shown to shape engagement and identity construction, whether through virtual doppelgangers~\cite{aymerich2014use, penn2010virtual} or interactions with past-self representations that support introspection and emotional processing~\cite{huang2022innerchild, misawa2015wearing}. Recent work outside of language education has started to investigate how AI self-clones may function as role models. As one example, incorporating AI self-clones as role models in presentation training has been found to improve learner experience alongside emotional and cognitive skill development~\cite{zheng2025learning}. 

Building on these strands of research, we frame AI self-clones not as direct replicas of a learner's current self but as aspirational representations---``better me'' versions that embody the learner's Ideal L2 Self. Rather than relying on social comparison with similar others~\cite{festinger1954theory}, this approach emphasizes the aspirational dimension of envisioning oneself as a more confident language user. This reframing shifts the role of self-clones from mere replication toward supporting growth, extending prior research on digital doubles into the context of language learning.

\section{System Design}
In this section, we introduce \emph{AI Twin}, a system designed for ESL learners to embody a ``better me'' by allowing learners to hear their own voice producing more fluent English speech. Rather than positioning feedback as external correction, the system reframes it through personalized self-representation, aiming to support both language development and learner motivation. We first describe why voice was chosen as the primary medium of self-representation, then outline the user flow that structures a typical practice session, and finally detail the system's architecture and implementation.

\subsection{Voice as medium of self-representation}
We chose voice as the primary medium for realizing the AI Twin. GenAI offers multiple ways of constructing self-representations---ranging from simulated personalities and thought patterns to visual embodiments such as avatars or deepfakes. While these modalities can shape identity cues in meaningful ways, our context required a medium that would both align with the learner's Ideal L2 Self and remain practical for ESL conversational practice.

Visual representations, for example, risk entering the ``uncanny valley'', which may distract learners or reduce authenticity in practice. By contrast, AI Twin's use of voice provides a more natural and immediate form of self-projection in spoken dialogue. According to Dörnyei's Ideal L2 Self framework, ``the more vivid and elaborate the possible self, the more motivationally effective it is expected to be.''
By hearing their utterances rephrased in a more fluent version of their own voice, learners could vividly imagine their aspirational self as a confident speaker, gaining the concreteness needed for motivation without the abstraction or artificiality associated with other modalities.

Focusing on voice also reflects the functional demands of ESL learning. Conversation is inherently auditory and temporal, and practicing with AI Twin can reinforce fluency goals by embedding feedback seamlessly within the act of speaking. Thus, voice served as both a theoretically grounded and practically effective medium for embodying the learner's idealized self in our design.

\begin{figure*}[!t]
    \centering
    \includegraphics[width=0.8\linewidth]{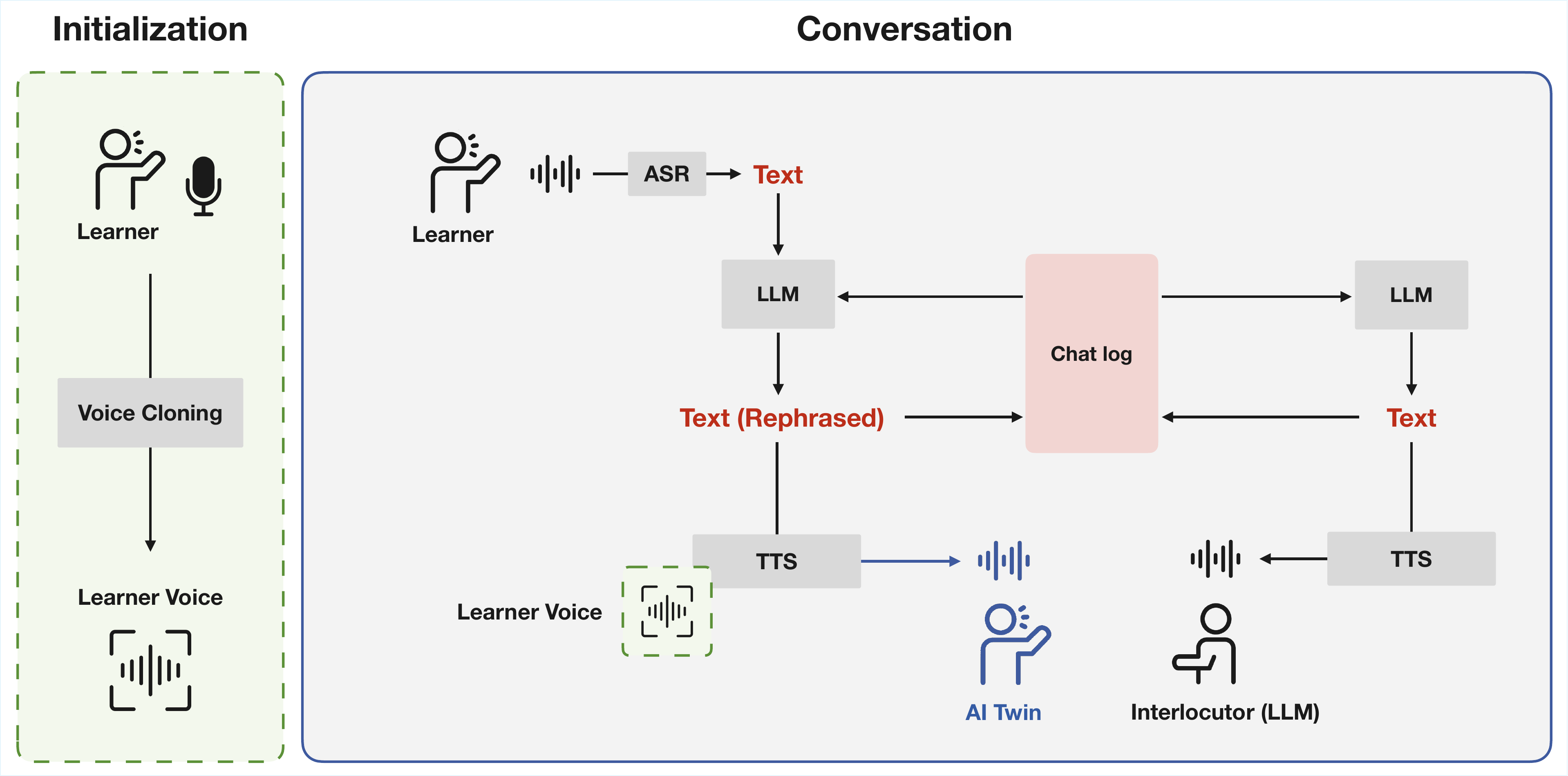}
    \caption{System flow of AI Twin. The process begins with initialization, where the learner registers their voice to create a personalized clone. During practice, the conversation cycle repeats: the learner's speech is transcribed using ASR, reformulated by a LLM, and synthesized in the learner's cloned voice. The AI interlocutor then generates a response based on this reformulated speech, enabling interactive conversation practice.}
    \Description{The figure shows the system flow of AI Twin from left to right. On the left, initialization involves the learner registering their voice to create a personalized clone. On the right, the conversation cycle illustrates the recurring process: the learner's spoken input is transcribed with ASR, reformulated by an LLM, synthesized with the cloned voice, and then used by the AI interlocutor to generate a reply, completing the interaction loop.}
    \label{fig:system_flow}
\end{figure*}

\subsection{User Flow}
\label{sec:subsection_userflow}
We designed AI Twin as a speech-based conversational interface in which all interactions occur through spoken language. Learners practice conversation with an \emph{AI interlocutor}, a role-playing conversational agent that serves as their practice partner. 

The session begins with voice registration: learners record a short passage (approximately 30 seconds), which is used to generate the AI Twin's synthetic voice. Once registered, learners begin a conversation session with an AI interlocutor. To participate, they must speak their responses aloud, simulating the conditions of natural face-to-face communication. We opted for a turn-based conversational flow, rather than real-time exchanges with potential interruptions, to provide learners a supportive space for speech production. This design emphasizes oral proficiency and comprehension, ensuring that spoken language remains the sole medium of interaction.

During each turn, the learner responds to the AI interlocutor. Before the interlocutor replies, AI Twin rephrases the learner's utterance into a more fluent version and delivers it in the learner's synthetic voice. The AI interlocutor then continues the conversation based on this rephrased input. This design embeds feedback directly into the conversational flow, giving learners exposure to more proficient language use without overt correction. For additional practice, learners can replay sentences to focus on listening and reinforce the rephrased utterances. Taken together, this user flow operationalizes the Ideal L2 Self by allowing learners to repeatedly experience and project a more confident, fluent version of themselves in conversation.

\subsection{System Architecture and Implementation}

Whereas Section~\ref{sec:subsection_userflow} outlined the learner-facing interaction flow, here we describe the system architecture that enables this experience. The overall pipeline is depicted in Figure~\ref{fig:system_flow}, and consists of an initialization step followed by a repeating conversational cycle.  

\subsubsection{Initialization}

During initialization, the system generates a personalized voice model from a short speech sample provided by the learner. Several few-shot voice cloning models were tested during early development, and we selected ElevenLabs~\cite{elevenlabs} based on pilot feedback indicating that its output was both clear and recognizable as the learner’s own voice. We implemented this using its instant voice cloning feature, which can generate a voice model from a short sample. This lightweight process enabled rapid onboarding and allowed learners to immediately interact with their AI Twin speaking in a voice similar to their own.

Modeling a plausible, more fluent version of the learner requires subtle control over timbre, prosody, and accent. In early prototyping, we explored an alternative pipeline in which fluent English speech was synthesized in a native speaker's voice and then converted to the learner's timbre. Although this preserved timbre, learners often judged the output as ``too fluent'' and difficult to recognize as their own voice. Instant voice cloning produced output that sounded more like the learner, preserving their timbre and general prosody. Yet when learners provided the speech sample in English, the clone often carried over their developing L2 accent, and the lightweight cloning method offered limited control to refine it. 

To address this variability in voice quality, we asked learners to record their initial sample in their first language. This allowed the clone to capture their authentic timbre and prosody, while the synthesis module generated reformulated utterances with a near-standard English accent. The resulting voice blended the learner's recognizable vocal identity with a subtly more fluent delivery. We also tuned generation parameters (\texttt{speed = 0.9} and \texttt{stability = 0.85}) to prevent the synthesized voice from diverging too far from the user's own.
While this process reduced variability compared to earlier prototypes, some perceptual variation remained---specifically, in how fluent the clone should sound and how similar it should be to the learner---and this variation was not quantitatively controlled. This design choice reflects the broader challenge of finding a \textit{happy medium}---speech that remains identifiable as the learner yet fluent enough to serve as an effective linguistic model---discussed further in Section~\ref{subsection_self_vs_rolemodel}.

\begin{figure*}[t]
    \centering
    \includegraphics[width=1\linewidth]{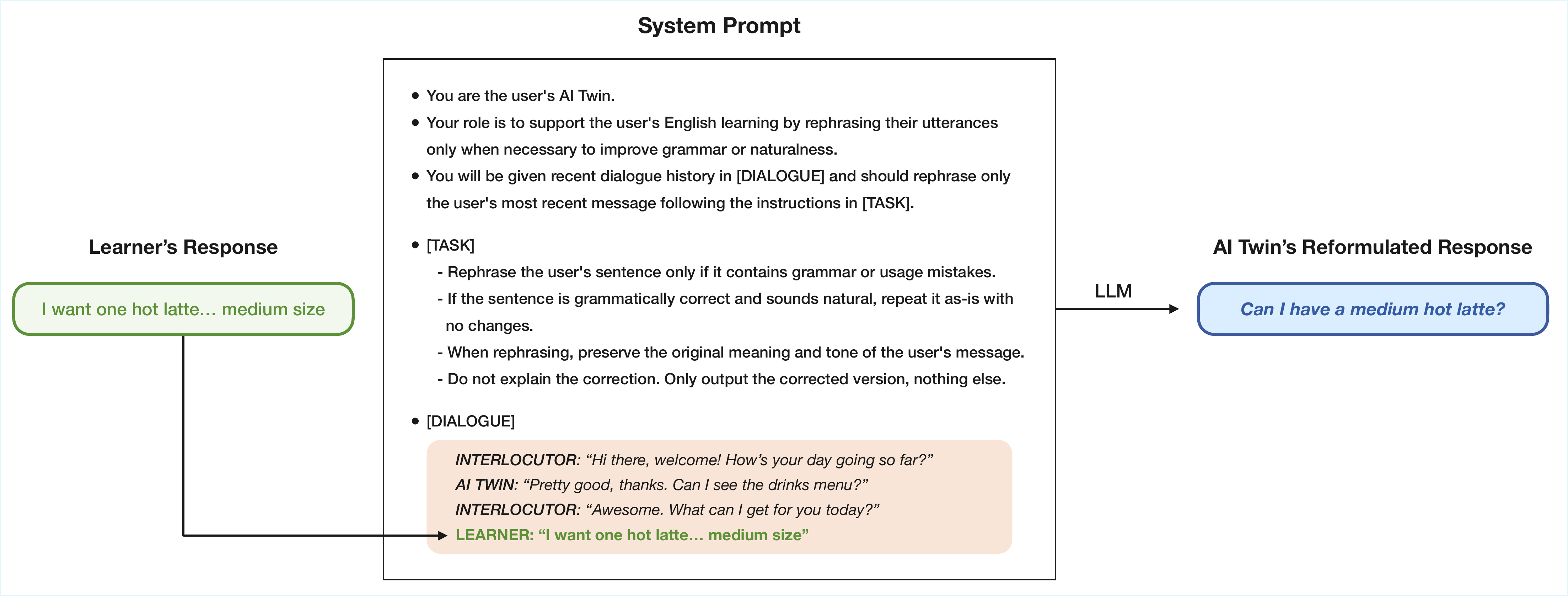}
    \caption{Rephrasing process in AI Twin. The learner's utterance, together with the ongoing dialogue context, is passed into the system prompt for the large language model (LLM). The LLM then generates a rephrased version of the learner's response, which is returned as clearer, more fluent English.}
    \Description{The figure illustrates the rephrasing process in AI Twin. The learner's spoken utterance, along with the dialogue context, is input into the system prompt for the LLM. The model outputs a rephrased version of the utterance, expressed in clearer and more fluent English, which is then returned to the learner.}
    \label{fig:ai_twin_prompt}
\end{figure*}

\subsubsection{Conversational Cycle}
For every user utterance, the system first applies automatic speech recognition (ASR) module (ElevenLabs \texttt{scribe v1}~\cite{elevenlabs}) to capture and transcribe the learner's spoken input. The transcription is then processed by the AI Twin module, which uses \texttt{gpt-4.1-mini} (OpenAI~\cite{openai}) to reformulate the utterance into a more fluent version while preserving its intended meaning, guided by the full dialogue context. The rephrased text is synthesized in the personalized voice model and passed to the AI interlocutor. 

The AI interlocutor then generates a response via \texttt{gpt-4.1-mini} (OpenAI~\cite{openai}) based on the reformulated utterance, rather than the learner's original transcription.
By grounding its output in the fluent version of the input, the system ensures that subsequent dialogue progresses smoothly and coherently, without being disrupted by grammatical errors or disfluencies. 
The interlocutor's textual reply is then rendered into natural speech through ElevenLabs~\cite{elevenlabs} text-to-speech (TTS) module, completing the conversational loop. 
This spoken output is returned to the learner, allowing them to engage in a continuous, voice-based interaction that mirrors the dynamics of human-to-human conversation.

This architecture maintains the natural flow of dialogue by embedding corrective feedback directly within the learner's turns. Instead of interrupting with overt error markings or explicit explanations, the system subtly integrates reformulated utterances as if they were the learner's own. 
In doing so, corrections function less like external interventions and more like self-modeled enhancements, reinforcing accurate forms while minimizing disruption to communicative intent.

\subsubsection{Prompt Design}
To guide the LLM in generating rephrasing, we designed prompts that emphasized \textit{minimal correction}. The model was instructed to preserve the learner's intended meaning, apply only the necessary grammatical or lexical adjustments, and avoid rewriting content. Providing the full dialogue history further supported contextual accuracy. The rephrasing process is illustrated in Figure~\ref{fig:ai_twin_prompt}, and specific prompt examples are provided in Appendix~\ref{appendix:prompts}.

\section{Evaluation}
We conducted a within-subject, mixed-method study to evaluate AI Twin against alternative feedback strategies. This design integrated quantitative measures of learner engagement with qualitative reflections, providing both systematic assessment and deeper insight into motivational effects. Because our primary goal was to understand learners' in-the-moment engagement and affective responses, the evaluation did not aim to assess learning gains, which typically require sustained, repeated practice or longitudinal observation~\cite{dekeyser2007skill}.
Consistent with Krashen and Terrell's \emph{Natural Approach} method, which suggests that natural interaction can lower learners' affective filters~\cite{krashen1983natural}, we did not impose a specific knowledge targets.
Instead, sessions were designed to be open-ended and communication-focused, encouraging spontaneous language use and reducing emotional barriers that might distort early engagement. This allowed the influence of each feedback strategy on natural language use and self-directed engagement to be more clearly observed.
Within this overall structure, the study focuses on how the three feedback strategies influence immediate engagement, perceived support, and motivational experience within these conversational tasks.
In the following sections, we describe the study details.

\subsection{Study Design and Conditions} 

\begin{figure*}
    \centering
    \includegraphics[width=0.8\linewidth]{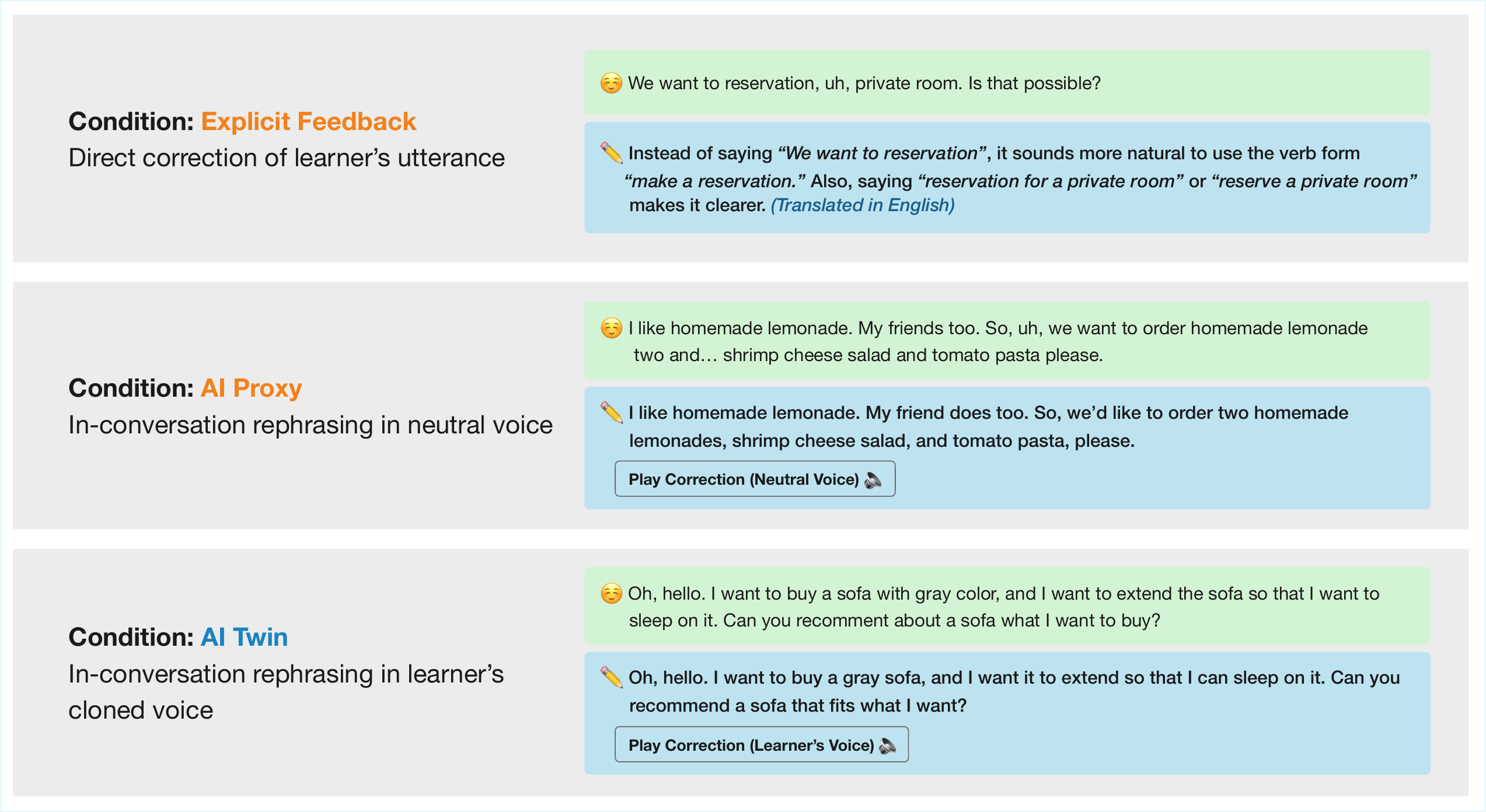}
    \caption{Examples of the three study conditions. After each learner utterance, (1) \emph{Explicit Feedback} provides direct correction, (2) \emph{AI Proxy} offers rephrasing in a neutral synthetic voice, and (3) \emph{AI Twin} (our approach) delivered the rephrased utterance in the learner's own cloned voice. Samples shown are drawn from the study data.}
    \Description{The figure shows three rows, each representing one study condition. In the first row, Explicit Feedback provides a direct correction of the learner's utterance. In the second row, AI Proxy rephrases the utterance into clearer English but delivers it in a neutral synthetic voice. In the third row, AI Twin also rephrases the utterance but delivers it in the learner's own cloned voice. Each row includes a sample drawn from the study data.}
    \label{fig:conditions}
\end{figure*}

\begin{figure*}[h]
    \centering
    \includegraphics[width=1\linewidth]{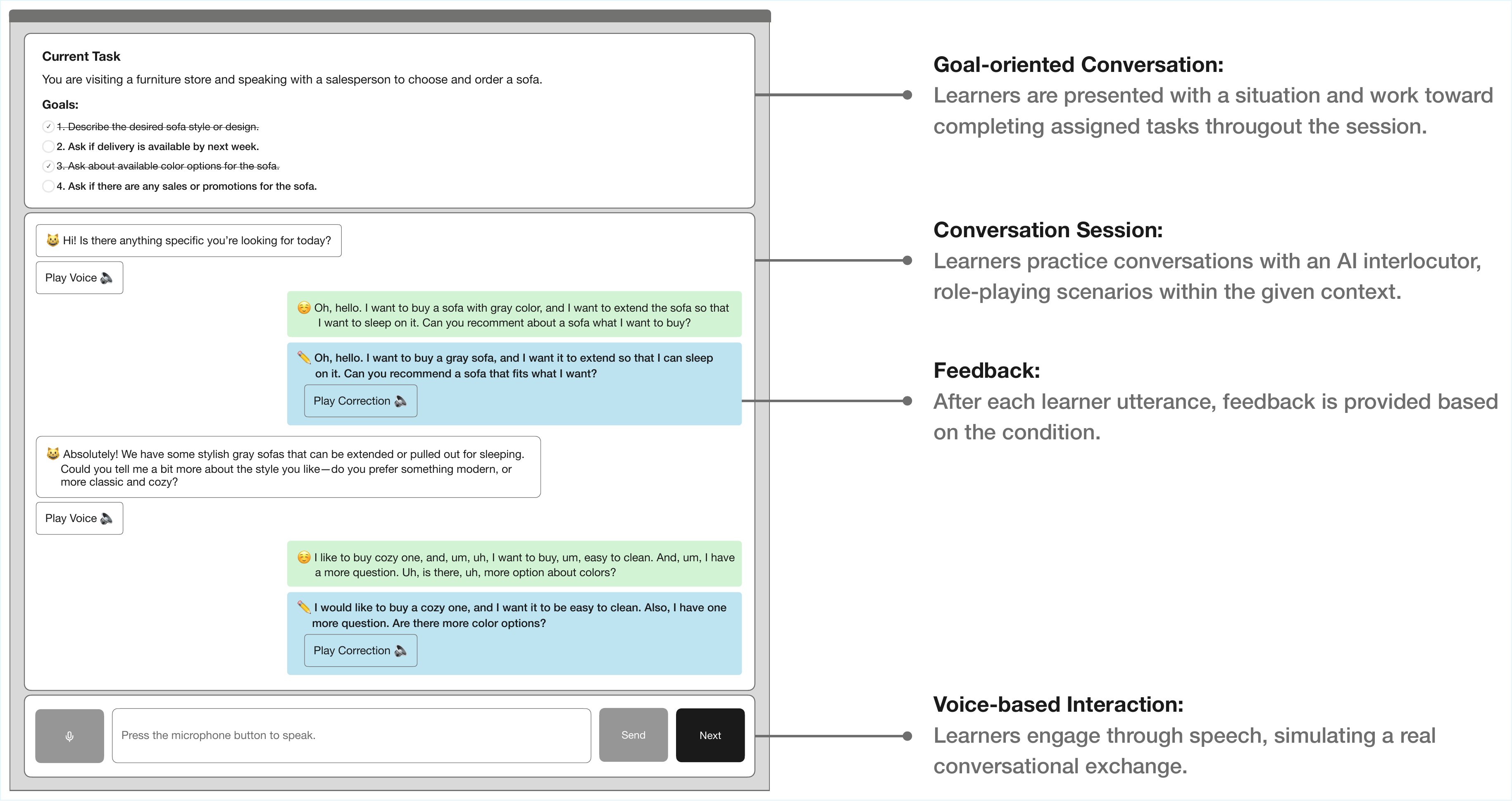}
    \caption{Illustration of the AI Twin interface used in the study. Learners engaged in voice-based interaction by recording their speech and listening to the system's responses. Within study sessions, practice was situated in goal-oriented conversations, and after each learner utterance AI Twin provided feedback through rephrased speech. The interface text was originally in Korean; English translations are shown here for clarity.}
    \Description{Illustration of the AI Twin interface used in the study. At the top, the current conversational task is displayed. Below, learners interact by pressing a microphone button to record their speech and then hear the system's spoken responses. The conversation is shown as text on screen, with the learner's original utterance and the system's rephrased version presented together.}
    \label{fig:screenshot}
\end{figure*}

\subsubsection{Baseline systems}
To situate the contribution of AI Twin, we compared it against two baseline conditions: (1) Explicit Feedback and (2) AI Proxy. These baselines were selected to distinguish the role of feedback style, contrasting rephrasing with direct correction, and the role of personalization, contrasting the learner's own voice with a neutral synthetic voice. Examples of each condition are shown in Figure~\ref{fig:conditions}.

\paragraph{\textbf{Baseline 1: Explicit Feedback}}
In this condition, instead of delivering the rephrased utterance in speech, the system presents explicit textual feedback highlighting grammatical errors, lexical alternatives, or improved expressions. 
This design reflects the dominant approach in existing AI-assisted ESL tools, enabling a direct comparison between implicit, conversationally embedded support and overt corrective feedback.

\paragraph{\textbf{Baseline 2: AI Proxy}}
This condition resembled our system in structure but, instead of using the learner's cloned voice, rephrased utterances were delivered in a non-personalized synthetic voice. Through pilot testing, we selected a neutral and non-distracting option---a female American-accented voice---so that the condition would provide a consistent proxy without strong stylistic cues~\cite{elevenlabs}.
This baseline allowed us to examine whether motivational effects arose specifically from hearing one's own personalized voice, or whether exposure to fluent rephrasings in a neutral proxy voice could provide a comparable benefit.

\subsubsection{Conversational Task Design}
Study sessions were built around goal-oriented conversational tasks that framed each interaction as a role-play with four short action items. Keeping tasks brief helped maintain participant attention and reduce the potential for carry-over effects across conditions in this within-subject design~\cite{greenwald1976within}.
Figure~\ref{fig:screenshot} illustrates the interface through which participants engaged in the conversational tasks during the study.

Rather than asking learners to initiate topics on their own or respond to broad, open-ended questions, each session was scaffolded around specific goals that guided the conversation.
This design offered two main advantages. 
First, it allowed for a more controlled conversational flow, ensuring that participants across conditions encountered comparable challenges and opportunities for practice. 
Second, it reduced the burden on learners, who did not need to generate topics or sustain small talk on their own. By framing each exchange around clear, immediate action items, we enabled learners to focus on producing language in context, while minimizing hesitation or anxiety over ``what to say next.'' 
In this way, goal-oriented conversation provided both methodological consistency and a supportive experience for participants.
Appendix~\ref{appendix:tasks} lists the detailed conversational tasks.

\subsection{Participants}
We recruited 20 adult South Korean ESL learners (ages 24--36, $M$=31.1, $SD$=3.44) who reported an interest in improving their spoken proficiency.  
Participants were recruited through online channels, as well as snowball sampling, which helped reach a diverse pool of learners.
All participants had completed at least secondary education and met the eligibility criteria of being able to read and speak in English within a web-based environment. 

We focused on adults rather than younger learners for two reasons. 
First, adults typically possess sustained exposure to English education in formal settings, which ensured a baseline of linguistic competence sufficient for engaging with our system. 
Second, adult learners are more likely to hold stable conceptions of learning and self-identity, making them an appropriate population for investigating the motivational dynamics of self-clone personalization.

\begin{figure}
    \centering
    \includegraphics[width=0.6\linewidth]{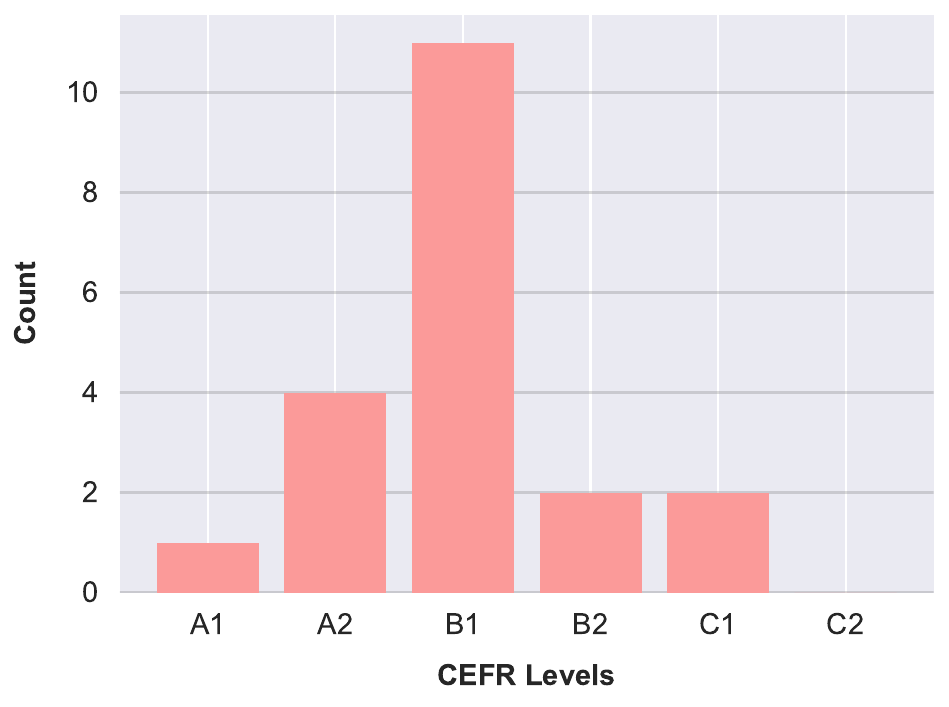}
    \caption{Distribution of participants' CEFR proficiency levels (N=20)}
    \Description{Histogram showing the number of participants at each CEFR level (A1-C2).}
    \label{fig:cefr}
\end{figure}

To characterize learners' language backgrounds, we collected participants' self-reported proficiency using the Common European Framework of Reference for Languages (CEFR) Self-Assessment Grid for the Spoken Interaction category, presented in a Korean-translated format for clarity~\cite{council2020cefr}. Reported levels ranged from A1 to C1, with most participants falling within the intermediate range (Figure~\ref{fig:cefr}).

\subsection{Study Procedure}
Each participant experienced all three feedback conditions in randomized order, following a within-subjects design. The study was conducted in person and lasted approximately one hour per participant. Before beginning, participants received an overview of the study procedures and data management practices and provided informed consent. 

After the briefing, each study session consisted of voice registration, three practice sessions with post-task surveys, and a concluding semi-structured interview. To initialize the voice-cloning system during voice registration, participants recorded a short speech sample by reading an English prompt aloud at a natural pace. They then completed three sessions of conversational interaction, each corresponding to one of the three feedback conditions. After each session, they filled out a brief questionnaire to capture immediate reactions.

The session concluded with a semi-structured interview guided by a fixed set of questions. The interview probed perceptions of the three feedback conditions with respect to emotional, cognitive, and behavioral engagement, directly aligning with our research questions. Participants were asked to reflect on how each condition influenced their confidence, sense of learning, and immersion in conversation. Combining questionnaire responses and interview accounts enabled triangulation, providing both broad patterns and nuanced perspectives on learner experience. Participants were compensated with a gift voucher equivalent to 30,000 KRW (approx. 22 USD). for their time. All procedures, including recruitment, consent, and compensation, were reviewed and approved by the Behavioural Research Ethics Board.

\subsection{Data Collection and Analysis}
Our evaluation focused on learner engagement, a construct widely described as comprising emotional, cognitive, and behavioral dimensions~\cite{kuh2009student}. To examine these dimensions, we collected quantitative data from questionnaires and qualitative data from semi-structured interviews, and applied complementary analyses. Questionnaire scores were compared across conditions using statistical tests, while interview accounts were analyzed thematically to explain or elaborate on the quantitative results. Together, these methods provided both systematic evidence and deeper insight into how different feedback styles shaped learner engagement. The following subsections detail the quantitative and qualitative components.

\subsubsection{Questionnaire Data}
To assess learner engagement quantitatively, we designed a questionnaire by adapting items from established instruments. Following prior work that conceptualizes engagement as comprising emotional, cognitive, and behavioral dimensions~\cite{fredricks2004school}, we selected and grouped items to reflect these three facets in our context.

\paragraph{Emotional Engagement.}
Emotional engagement was assessed using items from the Instructional Materials Motivation Survey (IMMS)~\cite{loorbach2015validation} and the Foreign Language Classroom Anxiety Scale (FLCAS)~\cite{horwitz1986foreign}.
From the IMMS, we included items from the Attention and Satisfaction subscales because they index affective responses such as interest, curiosity, enjoyment, and positive feelings toward the interaction. These constructs align with widely accepted definitions of emotional engagement as learners' emotional responses to a learning task~\cite{fredricks2004school}, and reflect the affect-based mechanisms underlying this component. We also incorporated items from the FLCAS to capture anxiety, a core emotional construct in language learning environments and a well-established marker of emotional engagement. Together, these measures allow us to characterize both positive and negative affective experiences during the session.

\paragraph{Cognitive Engagement.}
Cognitive engagement reflects the mental effort, strategy use, and self-regulation learners invest in the task. We used items from Greene et al.’s Original Cognitive Engagement (OCE) scale~\cite{greene2015measuring}, which examines strategic processing and sustained mental effort during learning activities. To capture additional aspects of learners' strategic processing and effort, we included items from the Motivated Strategies for Learning Questionnaire (MSLQ) self-efficacy component~\cite{pintrich1991manual}. Self-efficacy is widely recognized as a central predictor of cognitive engagement, influencing learners' strategic processing and persistence~\cite{greene2015measuring}, and has been shown to correlate positively with cognitive engagement in classroom learning~\cite{pintrich1990motivational}.
This categorization is consistent with prior work that applies the MSLQ as a measure of cognitive engagement, given its origins in assessing strategy use and self-regulation~\cite{pintrich1990motivational, fredricks2012measurement}.

\paragraph{Behavioral Engagement.}
Behavioral engagement refers to learners' observable effort, persistence, and involvement during the task. We used items from the Behavioral Engagement (BE) and Behavioral Disaffection (BD) subscales of the Engagement versus Disaffection with Learning scale (EvsD)~\cite{skinner2009motivational}, which assess attentional focus, effortful participation, and signs of withdrawal or disengagement. These items capture concrete behavioral indicators that can vary within a single session and provide a complementary perspective to the emotional and cognitive measures.

\hfill \break
All items were adapted to reflect interactions with the AI system rather than traditional classroom teachers (e.g., ``my language teacher'' was changed to ``AI system''). Participants responded on a 6-point Likert scale to encourage directional judgments and avoid neutral responding.
Reverse-coded items were adjusted prior to analysis, and subscale scores were computed by averaging item responses within each engagement category. 
To examine differences across the three feedback conditions, we conducted repeated-measures ANOVAs for each engagement dimension, followed by post-hoc paired $t$-tests to identify specific condition-level contrasts.
Moreover, to provide finer-grained insights beyond averaged subscale scores, we also conducted ANOVAs on each question to examine how responses varied across conditions.
Details of the questionnaire items are provided in Table~\ref{tab:per_question}.

\subsubsection{Interview Data}
Semi-structured interviews complemented the questionnaire by exploring learners' subjective experiences across the three engagement dimensions. Participants were first asked to share their overall impressions of the three conditions, and then to compare them. They reflected on differences between implicit rephrasing and explicit correction (AI Proxy, AI Twin vs. Explicit Feedback), as well as between feedback in their own voice and in a neutral synthetic voice (AI Twin vs. AI Proxy). This qualitative component enabled learners to articulate not only how they felt and behaved in each condition, but also how different feedback styles shaped their engagement.

For analysis, interviews were audio-recorded, transcribed verbatim, anonymized, and then translated into English. During translation, we added condition labels for clarity (e.g., ``first session'' → ``AI Twin condition''), since session order had been randomized.
Analysis followed a primarily deductive approach: responses were grouped according to the engagement dimensions and comparisons across different conditions. This structured focus allowed us to directly align participants' qualitative reflections with the quantitative survey results, while still capturing illustrative nuances in how learners experienced the different conditions.
Analysis therefore followed a primarily deductive approach: responses were grouped according to the engagement dimensions and feedback comparisons (implicit vs. explicit; personalized vs. non-personalized). This structured focus allowed us to directly align participants' qualitative reflections with the quantitative survey results, while still capturing illustrative nuances in how learners experienced the different conditions.

\renewcommand{\arraystretch}{1.5}
\begin{table*}[ht]
\small
\centering
\resizebox{0.7\linewidth}{!}{%
\begin{tabular}{cccc ccc}
\hline
\textbf{Measure} & \textbf{Explicit Feedback} & \textbf{AI Proxy} & \textbf{AI Twin} & \textbf{F(df$_1$, df$_2$)} & \textbf{$p$} & \textbf{$\eta_p^2$} \\ \hline
\textbf{Emotional Engagement}  & 4.03 (1.12) & 4.77 (0.67) & 4.83 (0.63) & F(2, 38) = 10.89 & .0002* & 0.36 \\
\textbf{Cognitive Engagement}  & 4.69 (1.01) & 4.66 (0.80) & 4.79 (0.52) & F(2, 38) = 0.29 & .75    & 0.02 \\
\textbf{Behavioral Engagement} & 4.48 (1.02) & 4.55 (0.83) & 4.60 (0.67) & F(2, 38) = 0.35 & .71    & 0.02 \\ \hline
\end{tabular}}
\vspace{5pt}
\caption{Means and standard deviations of engagement measures across conditions, with results of repeated measures ANOVA. Values in condition columns are reported as mean (SD).}
\label{tab:anova}
\end{table*}

\section{Findings}

Our evaluation of AI Twin revealed that it played a particularly strong role in fostering learners' emotional engagement.  Rephrasing in general supported engagement, while AI Twin further amplified this effect. Participants noted that interacting with their AI self-clone supported motivation, reduced speaking anxiety, and created a more enjoyable conversational experience. In what follows, we detail these findings, highlighting how different feedback conditions shaped learners' emotional, cognitive, and behavioral engagement with the system.

\subsection{How Rephrasing Supported Learner Engagement}
\subsubsection{Emotional Engagement}
\begin{figure}
    \centering
    \includegraphics[width=1\linewidth]{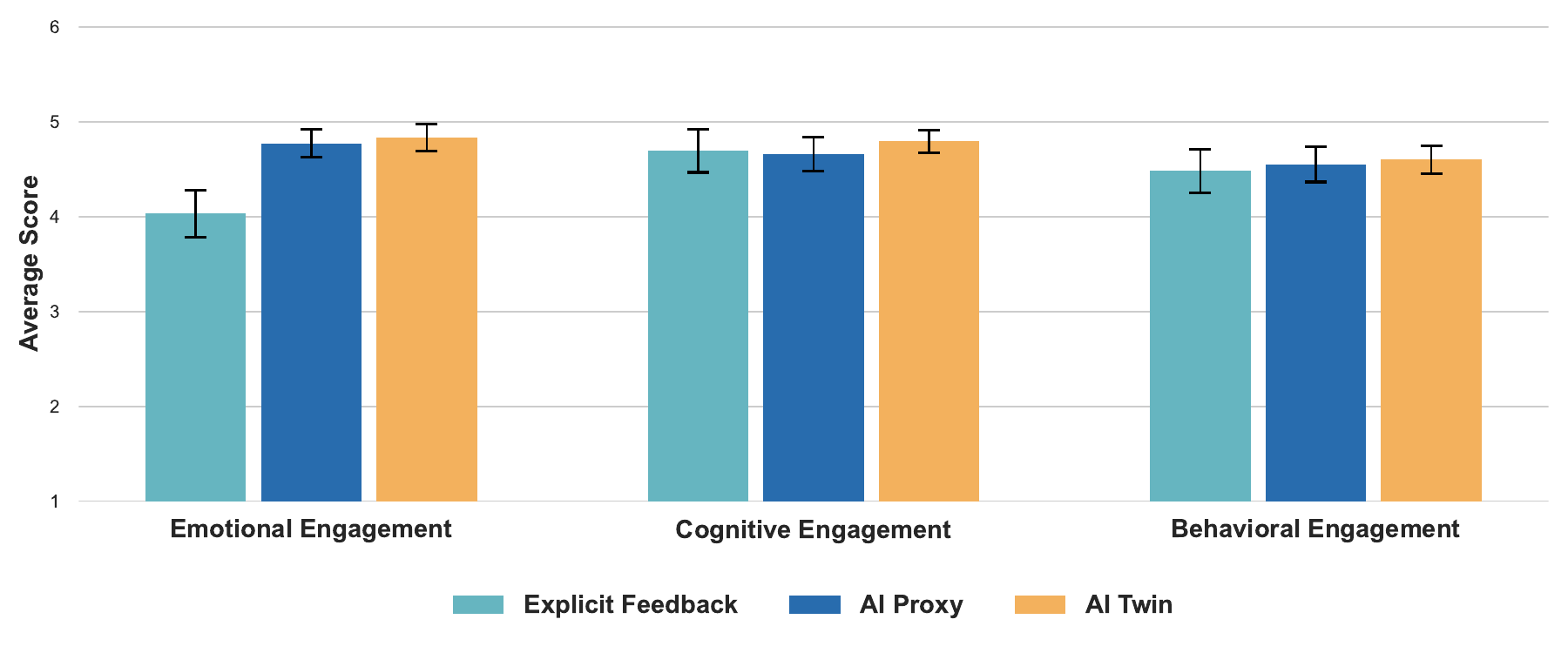}
    \caption{Average scores of emotional, cognitive, and behavioral engagement across conditions on a 6-point Likert scale. Error bars represent standard errors.}
    \Description{Bar chart showing average engagement scores for the three dimensions---emotional, cognitive, and behavioral---across the study conditions. Scores are plotted on a 6-point Likert scale, with error bars indicating standard errors. The exact means and standard deviations for each condition are reported separately in Table~\ref{tab:anova}.}
    \label{fig:score_plot}
\end{figure}

We first observed the general effectiveness of rephrasing (AI Proxy and AI Twin) as a feedback strategy. Quantitative results showed that rephrasing, whether personalized or not, significantly enhanced emotional engagement compared to Explicit Feedback condition. A repeated measures ANOVA revealed a significant effect of condition ($F(2, 38) = 10.89$, $p < .001$). Post-hoc comparisons indicated that both rephrasing conditions yielded higher emotional engagement than explicit correction (AI Proxy vs. Explicit Feedback: $t(19) = 3.73$, $p = .001$; AI Twin vs. Explicit Feedback: $t(19) = -3.46$, $p = .003$), with no difference between the two rephrasing formats ($t(19) = -0.48$, $p = .64$). This suggests that the conversational rephrasing format itself played an important role in improving emotional engagement.
The overall comparison across conditions and engagement types is illustrated in Figure~\ref{fig:score_plot}. Detailed repeated measures ANOVA results are reported in Table~\ref{tab:anova}, and post-hoc paired $t$-test results for emotional engagement are presented in Table~\ref{tab:post_hoc}.

Per-question analyses further showed that most emotional engagement items differed significantly across conditions, with the exception of several nervousness-related items (e.g., ``I felt nervous when I had to speak during this session'', ``I got so nervous I forgot things I know.'', ``I got nervous when the AI asked questions which I haven't prepared in advance.''). To clarify the direction of effects, paired $t$-tests on significantly different items revealed a consistent pattern: for the majority of items (Items 1, 2, 3, 6, 8, 11, and 12), scores in the Explicit Feedback condition were significantly lower than in both rephrasing conditions. These results reinforce that embedding corrective feedback within the conversational flow helps foster more positive emotional experiences.
Full per-question statistics are provided in Table~\ref{tab:per_question}, and detailed post-hoc results for the items that showed significant condition differences are provided in Appendix~\ref{appendix:per_question_post_hoc}.

\renewcommand{\arraystretch}{1.5}
\begin{table}
\small
\centering
\begin{tabular}{cclc}
\hline
\textbf{Comparison} & \textbf{$t$} & \textbf{$p$} & \textbf{$d_z$} \\ \hline
\textbf{AI Proxy vs Explicit Feedback} & 3.73 & .001* & 0.83 \\
\textbf{AI Proxy vs AI Twin} & -0.48 & .64 & -0.11\\
\textbf{Explicit Feedback vs AI Twin} & -3.46 & .003* & -0.77\\ \hline
\end{tabular}
\vspace{5pt}
\caption{Pairwise post hoc comparisons between conditions for emotional engagement (paired $t$-tests). Reported values are the test statistic ($t$), $p$-value, and within-subjects effect size ($d_z$).}
\label{tab:post_hoc}
\end{table}

\renewcommand{\arraystretch}{1.3}
\begin{table*}[t]
\centering
\scriptsize
\resizebox{\textwidth}{!}{%
\begin{tabular}{c m{6cm} cccc}
\hline
\rowcolor{gray!30}
\multicolumn{1}{c}{\#} &
\multicolumn{1}{c}{\textbf{Item}} &
\multicolumn{1}{c}{\textbf{Explicit Feedback}} &
\multicolumn{1}{c}{\textbf{AI Proxy}} &
\multicolumn{1}{c}{\textbf{AI Twin}} &
\multicolumn{1}{c}{\textbf{p}} \\ \hline
\rowcolor{gray!15}
\multicolumn{6}{l}{\textbf{Emotional Engagement}} \\ \hline
1 & Completing the exercises in this lesson gave me a satisfying feeling of accomplishment. (IMMS, Item 05S01) & 4.15 (1.46) & 4.85 (1.18) & 5.05 (1.19) & \textbf{.007**} \\
2 & I enjoyed this lesson so much that I would like to know more about this topic. (IMMS, Item 14S02) & 3.95 (1.23) & 4.7 (1.13) & 5.05 (1.28) & \textbf{.002**} \\
3 & I really enjoyed studying this lesson. (IMMS, Item 21S03) & 4.0 (1.41) & 4.85 (1.18) & 5.05 (0.89) & \textbf{.0007***} \\
4 & The way the feedback is presented in the sessions helped keep my attention. (IMMS, Item 17A06) & 4.05 (1.61) & 4.4 (1.27) & 4.85 (0.93) & .11 \\
5 & The session was not engaging and unappealing. (IMMS, Item 15A05)  & 4.45 (1.5) & 4.9 (1.21) & 5.35 (0.59) & \textbf{.02*} \\
6 & I didn't worry about making mistakes while using this learning system. (FLCAS, Item 2)  & 3.85 (1.69) & 5.1 (0.91) & 4.85 (0.99) & \textbf{.0002***} \\
7 & I felt nervous when I had to speak during this session. (FLCAS, Item 3)  & 3.65 (1.5) & 4.3 (1.3) & 4.15 (1.27) & .08 \\
8 & It wouldn't bother me at all to do more. (FLCAS, Item 5)  & 4.5 (1.0) & 5.1 (1.02)  & 5.1 (0.85)  & \textbf{.01*} \\
9 & I got so nervous I forgot things I know. (FLCAS, Item 12)  & 4.2 (1.64) & 4.65 (1.18) & 4.75 (1.33) & 0.30 \\
10 & I felt confident using English during this session. (FLCAS, Item 18) & 3.8 (1.2) & 4.65 (1.18)  & 4.3 (1.22)  & \textbf{.008**} \\
11 & I was afraid that the system is ready to correct every mistake I made. (FLCAS, Item 19)  & 3.65 (1.81) & 5.2 (1.11) & 5.1 (1.21)  & \textbf{.00003***} \\
12 & I don't feel pressure to prepare very well when I responded. (FLCAS, Item 22) & 3.85 (1.66) & 4.95 (1.19) & 4.9 (1.12)  & \textbf{.008**} \\
13 & I got nervous when the AI asked questions which I haven't prepared in advance. (FLCAS, Item 33) & 4.3 (1.34) & 4.4 (1.35) & 4.35 (1.46) & .95 \\ \hline
\rowcolor{gray!15}
\multicolumn{6}{l}{\textbf{Cognitive Engagement}} \\ \hline
14 & I am confident I can talk about [topic] in a short English conversation. (MSLQ, Item 12) & 4.35 (1.27) & 4.7 (1.03) & 4.6 (0.94) & .21 \\
15 & I am certain I can use simple English phrases and sentences to describe [topic]. (MSLQ, Item 29) & 4.45 (1.1) & 4.85 (0.88) & 4.9 (0.72)  & .19   \\
16 & If I had to use this system or complete a similar task again, I expect to do well. (MSLQ, Item 21) & 4.9 (1.25) & 4.95 (1.05) & 5.1 (0.91)  & .75   \\
17 & While learning new concepts, I tried to think of implications and practical applications. (OCE, Item 21) & 4.45 (1.39) & 4.2 (1.24) & 4.3 (1.13)  & .75   \\
18 & I tried to organize the feedbacks in a way that made sense to me. (OCE, Item 22) & 4.7 (0.92) & 4.5 (1.1) & 4.65 (1.04) & .71 \\
19 & Within the session I made sure I understood the lesson content. (OCE, Item 26) & 4.8 (1.15) & 4.45 (1.15) & 4.7 (1.03)  & .41 \\
20 & I tried to check what my errors were during the session. (OCE, Item 28)  & 5.2 (1.11) & 4.95 (1.0) & 5.3 (0.8)   & .33 \\ \hline
\rowcolor{gray!15}
\multicolumn{6}{l}{\textbf{Behavioral Engagement}} \\ \hline
21 & I tried hard to do well. (EvsD, BE Item 1) & 4.6 (1.35) & 4.75 (1.41) & 4.8 (1.32)  & .53 \\
22 & I paid attention. (EvsD, BE Item 4) & 5.05 (1.0) & 5.35 (0.75) & 5.15 (0.75) & .20 \\
23 & I participated very carefully. (EvsD, BE Item 1) & 4.15 (1.35) & 3.75 (1.37) & 3.4 (1.39)  & \textbf{.04*} \\
24 & I thought about other things during the session. (EvsD, BD Item 4) & 4.8 (1.24)& 5.0 (1.03) & 5.15 (1.09) & .36 \\
25 & I did just enough to complete the given tasks. (EvsD, BD Item 3) & 3.8 (1.54) & 3.9 (1.59) & 4.5 (1.43)  & .10 \\ \hline
\end{tabular}
}
\vspace{5pt}
\caption{Engagement questionnaire items, with the corresponding source provided in parentheses, along with mean (SD) responses across the three feedback conditions. The final column reports per-question ANOVA $p$-values, with significance indicated using conventional thresholds (* $p<.05$, ** $p<.01$, *** $p<.001$).
}
\label{tab:per_question}
\end{table*}

Qualitative accounts deepened this finding by illustrating why rephrasing was experienced as more emotionally supportive. Participants often described explicit correction as discouraging, with one noting that when every mistake was listed, ``it really hurt my feelings. It felt like someone next to me was constantly saying, `No, that's wrong' '' (P5). In contrast, rephrasing was experienced as less judgmental and more supportive of natural conversation. As P10 reflected, ``In Explicit Feedback condition, I actually felt increasing pressure that I shouldn't make mistakes, whereas in AI Proxy and AI Twin conditions, I was able to speak more naturally and confidently.'' Together, these accounts suggest that by embedding corrections within the flow of conversation, rephrasing reduced anxiety and fostered confidence.

\subsubsection{Cognitive Engagement}
\label{sec:rephrase_cog}
For cognitive engagement, participants expressed divergent views regarding the effectiveness of rephrasing. Some regarded it as valuable for drawing attention to pronunciation and natural expression. As one participant observed, ``from the perspective of speaking, things like pronunciation and the actual voice are important. That's why I felt the spoken version [AI Proxy and AI Twin conditions] was better'' (P11).

However, others reported a stronger sense of learning with Explicit Feedback condition. P9 reflected, ``I think I learned the most from the Explicit Feedback condition. The feedback was detailed, whereas in the AI Proxy and AI Twin conditions there wasn't much explanation, so I just kind of accepted them and moved on.'' Similarly, some participants with lower proficiency expressed a preference for Explicit Feedback condition, as it provided explanations in their first language (P2). A further limitation identified was that rephrasing sometimes obscured the underlying error, as one participant explained: ``I could just go, ‘ah, okay,' and move on … but at the same time, I was a little confused about exactly what was wrong'' (P19). While not a strict recast, this observation echoes prior work showing that recasts can preserve conversational flow but leave learners uncertain about the correction~\cite{lyster1998negotiation, kim2004issues}.

Taken together, these perspectives suggest that while rephrasing supported attention to fluency and naturalness, explicit correction more often created a perception of learning, highlighting a trade-off between conversational flow and the clarity of instructional feedback.

\subsubsection{Behavioral Engagement}
Although quantitative results did not reveal statistically significant differences, participants' reflections pointed to distinctions in how feedback styles shaped immersion and attention. Rephrasing was often associated with a stronger sense of immersion and smoother conversational flow. One participant explained, ``I think I was most immersed in the situation during the AI Proxy condition. Instead of reading through each piece of feedback one by one, I just heard my response expressed in a more refined way and then immediately continued to the next turn. That really gave me the feeling of having a real conversation'' (P4).

In contrast, Explicit Feedback condition was described as disruptive to attention, as another participant noted: ``In the case of Explicit Feedback condition, the feedback contained so much text that it was a bit difficult to focus on the conversation while reading through it all'' (P10). These accounts suggest that embedding corrections within the conversational flow supported behavioral engagement by sustaining immersion, whereas explicit correction risked diverting attention away from the conversation.

Complementing these qualitative patterns, per-question analyses revealed that one behavioral item---``I participated very carefully''---showed a significant difference across conditions, with post-hoc paired $t$-tests indicating that scores in the Explicit Feedback condition were significantly higher than in the AI Twin condition (see Appendix~\ref{appendix:per_question_post_hoc}). This suggests that although explicit correction may disrupt immersion, it may simultaneously prompt more deliberate, effortful participation.

\begin{figure*}[ht]
    \centering
    \includegraphics[width=0.9\linewidth]{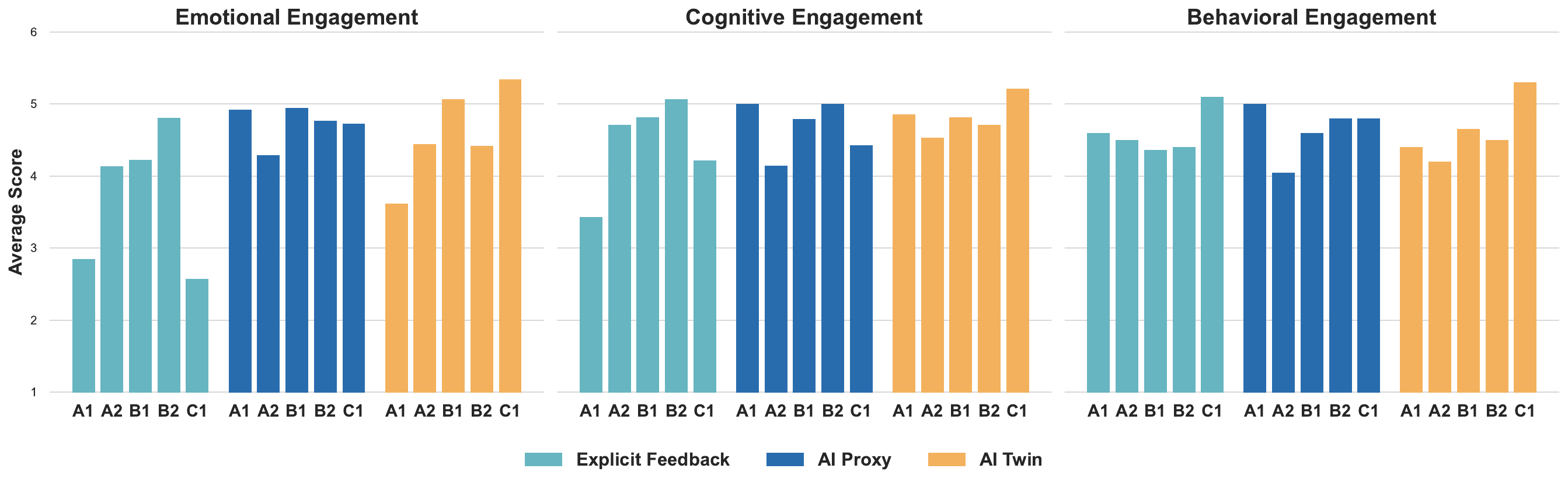}
    \caption{Average emotional, cognitive, and behavioral engagement scores across CEFR levels (A1--C1), grouped by feedback conditions}
    \Description{The figure presents three side-by-side bar plots showing average emotional, cognitive, and behavioral engagement (left to right). Within each plot, engagement is broken down by three feedback conditions (Explicit Feedback, AI Proxy, and AI Twin; left to right) and for each condition, five bars represent the mean scores for each CEFR level (A1–C1).}
    \label{fig:cefr_posthoc}
\end{figure*}

\subsection{AI Twin added value in positive engagement}
\label{subsection_findings_twin}
Although statistical tests revealed no significant differences between the two rephrasing conditions---AI Proxy and AI Twin, participants' accounts suggested that the AI Twin provided unique benefits across emotional, cognitive, and behavioral engagement.
To assess whether these benefits stem from first-encounter effects rather than the design itself, we conducted a targeted order analysis comparing participants who experienced AI Twin first ($N=11$) with those who first interacted with AI Proxy ($N=9$). Mixed ANOVAs revealed no significant \emph{Condition} $\times$ \emph{Order} interactions (emotional: $F(2,36)=0.84$, $p=.44$; cognitive: $F(2,36)=0.49$, $p=.61$; behavioral: $F(2,36)=1.25$, $p=.30$). While not definitive, this pattern offers no strong indication that encountering the voice-cloned system first substantially shaped the observed effects.

\subsubsection{Emotional Engagement}
The strongest added value of the AI Twin lay in emotional engagement. Part of this came from the experience of hearing one's own voice. Participants described it as enjoyable and novel. One explained, ``It created a sense of anticipation. I looked forward to the response [to hear my voice]'' (P5), while another added, ``It was more fun to learn this way'' (P12). The personalization of hearing their utterances echoed back in their own voice created a sense of curiosity and playfulness that made practice feel different from ordinary study.

Beyond the voice itself, the system design created a more positive and reassuring practice environment. Several participants explained that the AI Twin reduced pressure by making practice feel more familiar and private. The familiarity of hearing their own voice helped them relax, as one noted, ``Since it was my own voice, it felt familiar, so it helped me relax and ease my tension'' (P17). Others emphasized how the system created a sense of privacy, making the experience feel less exposed: ``It felt like I was the only one listening'' (P9). Together, this familiarity and privacy shaped a more forgiving environment, where mistakes felt less consequential. As P17 further reflected, ``I also felt a bit of a mindset that it was okay to make mistakes.''

Taken together, these reflections suggest that the AI Twin enriched emotional engagement both through the novelty of voice personalization and through a system design that fostered comfort, reduced anxiety, and supported immersive practice.

\subsubsection{Cognitive Engagement}
Beyond enjoyment, the AI Twin supported moments of reflection by revealing discrepancies between how learners perceived their speech and how it actually sounded. One participant described noticing a mismatch between their internal impression and the cloned playback: 

\begin{quote}
``When I heard it in my own voice, there was a bit of a gap between the voice I hear in my head and the cloned one. It made me realize, ‘Oh, so that's my intonation. It's not as lively as I thought. I thought I was speaking more energetically, but actually not.' I think I was able to catch those things more clearly.'' (P7)
\end{quote}

This account illustrates how the AI Twin prompted learners to recalibrate their self-perception, making subtle aspects such as intonation and expressiveness more visible. In this way, it facilitated moments of heightened self-awareness that could inform learners' understanding of their own speech patterns.

\subsubsection{Behavioral Engagement}
Participants expressed mixed experiences regarding behavioral engagement. For some, the AI Twin encouraged closer attention to language use. One participant explained, ``I wanted to try saying it that way. So I found myself paying closer attention to the changed words'' (P13). This suggests that hearing rephrases in their own voice could prompt learners to engage more actively with the corrections. At the same time, a few participants found the novelty distracting, admitting that they focused more on the voice than the content: ``When I listened to the rephrased sentences in my own cloned voice, I found myself focusing more on the voice than the content'' (P4). These contrasting accounts indicate that while the AI Twin sometimes deepened attention and practice, its effectiveness varied depending on individual preference.

\subsection{AI Twin enhanced motivation}
Participants often described how their positive engagement during practice translated into stronger motivation to continue learning. When asked to identify the most motivating condition, 13 of 20 participants selected AI Twin, underscoring its role in sustaining learners' commitment. Much of this motivational effect came from how the system aligned with learners' aspirational self-image as more fluent English speakers, supporting both their sense of self-efficacy and their Ideal L2 Self.

For some, hearing their rephrased utterances in their own voice fostered a sense of achievement and belief in their own ability to improve. One participant explained,
\begin{quote}
    ``More than anything, it made me feel a stronger sense of achievement. I thought, ‘If I just polish a little more, I could speak like this too.' Hearing a similar voice to mine gave me that feeling, so I think I'd be more motivated to actively use this app or study more.'' (P13)
\end{quote}
Another described how the experience created an impression of progress, even if imagined: receiving feedback in their own voice ``felt as if my English speaking skills had improved a bit,'' which gave them ``a kind of unwarranted confidence'' to keep trying (P14). These accounts show how the AI Twin strengthened learners' belief that their goals were attainable, a key component of self-efficacy.

Other participants placed less emphasis on what they could do at present, instead concentrating on the speaker they hoped to become. Hearing a more fluent version of themselves made learners aware of the gap between their current abilities and their desired identity as competent English speakers, which in turn motivated them to continue improving. As P16 explained, 
\begin{quote}
    ``Receiving feedback in my own voice gave me the strongest feeling of wanting to improve myself. [...] When I heard it in my own voice, I felt this emotion of `I want to speak well,' so I paid more attention to speaking with better expressions.'' (P16)
\end{quote}
This illustrates how hearing a more fluent version of themselves not only strengthened motivation but also directed greater attention to accuracy and expression during practice. These reflections capture the motivational pull of the Ideal L2 Self, where learners are driven by the desire to align with an imagined, more proficient identity as English speakers.

Overall, these perspectives illustrate two complementary motivational pathways: strengthened self-efficacy, reflected in a belief that improvement was within reach, and orientation toward the Ideal L2 Self, reflected in a desire to align with a more fluent version of oneself. By simultaneously reinforcing learners' confidence in their ability and their vision of who they could become, the AI Twin offered a uniquely motivating experience. At the same time, not all participants experienced the discrepancy positively. As one acknowledged, ``When I listened in my own voice, the gap between what I said and the AI's version was so clear that I felt a bit timid'' (P13). 
This highlights that while confronting the gap between one's current and aspirational self often fostered motivation, for some learners it could instead be momentarily discouraging.

\subsection{Preferences Reflected a Balance of Engagement and Learning}
While the AI Twin fostered strong engagement and motivation, not all participants felt it supported learning as effectively as Explicit Feedback condition.
As noted in Section ~\ref{sec:rephrase_cog}, several described experiencing a stronger ``feeling of learning'' when corrections were presented directly, and this perception influenced their preferences. 
When asked in our interviews which system best supported learning, participants were split: 9 of 20 participants judged Explicit Feedback condition to be more effective, while another 9 favored the AI Twin. Interestingly, these judgments did not always align with what they preferred to use in practice. 
When asked which system they would actually choose, 10 of 20 selected the AI Twin, while 6 favored Explicit Feedback condition.

These results suggest that preferences were shaped not by one factor alone but by how learners balanced engagement and perceived learning. Some prioritized the immediate clarity of direct correction---``For me, it's Explicit Feedback condition. I think a learning app should provide direct corrections'' (P11)---while others emphasized conversational ease---``If I were using it for studying, I'd probably choose the second session. On the other hand, if I just wanted to chat comfortably in English, I think I'd use the first session'' (P3). Still others expressed interest in hybrid designs that would allow them to shift between modes depending on their goals, such as one participant's suggestion for a toggle to reveal explicit feedback only when needed (P7).

Taken together, these reflections show that engagement and perceived learning were not strictly in opposition. Rather, learners weighed them differently, and their preferences reflected the balance they saw between enjoying practice and feeling that they were making progress. Within this balance, the AI Twin often emerged as the preferred option, even when Explicit Feedback condition was judged to better support learning.

\paragraph{Post-hoc analysis by language proficiency.}
Post-hoc descriptive analysis explored how learners' CEFR levels may shape their experience with the systems. Among basic users (A1–A2), only 1 of 5 preferred the AI Twin; the remaining four favored Explicit Feedback ($N=2$) and AI Proxy ($N=2$). Notably, both proficient users (C1) selected the AI Twin as their preferred option. To contextualize these differences, Figure~\ref{fig:cefr_posthoc} presents average emotional, cognitive, and behavioral engagement scores across proficiency levels for each feedback condition. 
Although based on a small number of high-proficiency participants ($N=2$), these exploratory descriptive patterns point to possible differences in how learners at different proficiency levels orient toward feedback formats, with higher-proficiency learners potentially being more receptive to the fluency-oriented interaction style of the AI Twin.

\section{Discussion} 
\label{section_discussion}
\subsection{Ideal L2 Self vs Role-Model}
\label{subsection_self_vs_rolemodel}
Although many participants responded positively to hearing their utterances rephrased in their own voice, others expressed a clear preference for non-personalized rephrasing (AI Proxy). This preference was often tied either to dissatisfaction with the sound of their own voice or to a stronger motivational pull toward a fluent, native-sounding voice. One participant explained, ``It might be because I generally don't like my own voice very much'' (P20). Others valued the proxy voice, noting that it ``sounded a bit more professional'' than their own (P17), or describing aspirational reactions such as, ``I envy her---she speaks English so well and confidently'' (P5).  

These accounts highlight that personalization is not uniformly beneficial; its effectiveness depends on learners' orientations toward self-representation and linguistic identity. Prior work similarly shows that learners often prefer synthesized voices with particular qualities, such as fluency or professionalism~\cite{orii2022designing, rubin2015capture}. Beyond these perceptual preferences, however, some participants framed the proxy voice as a \emph{role-model}---an aspirational figure embodying the qualities of proficient speech. This orientation contrasts with alignment to one's \emph{Ideal L2 Self}, instead positioning the system as an external model to emulate, echoing findings on the motivational power of role models in language learning~\cite{muir2021role}.  

Taken together, these findings suggest the importance of flexibility. Designing systems that can accommodate both self-referential and role-model orientations may be critical for supporting heterogeneous learner motivations.

\subsection{Differing Conceptions of Learning}
Our findings also revealed divergent learner conceptions of what constitutes ``real learning.'' 
Several participants expressed a preference for explicit corrective feedback, reasoning that overt grammatical correction provided clear action items and thus represented more tangible progress. 
This aligns with prior work in SLA showing that explicit correction, while potentially intrusive, is highly salient and effective in drawing learners' attention to the gap between their current performance and target norms~\cite{ellis2006implicit, oliver2003interactional}. 
By contrast, rephrasing was often perceived as less instructive, since it preserved conversational flow but obscured the precise nature of the error. 
These perspectives suggest that preferences for feedback styles are tied not only to affective responses but also to learners' epistemic beliefs about how language learning occurs.

From a design standpoint, these findings highlight the potential value of hybrid strategies that integrate explicit and implicit feedback. For instance, systems could maintain rephrasing during conversation to sustain immersion, while providing explicit corrective explanations after each session to support reflection and longer-term uptake.
Furthermore, the tension between fluency-oriented and form-focused feedback underscores the need for longitudinal efficacy studies. 
Such work would help clarify how different feedback modalities contribute to learning gains across proficiency levels, and inform the optimal balance between affective support and instructional clarity in AI-mediated language learning environments.

\subsection{Expansion to Second Language Acquisition}
While our study focused on English as a Second Language (ESL), the underlying mechanisms of AI Twin extend to second language acquisition (SLA) more broadly. 
Importantly, the framework of the Ideal L2 Self suggests that personalization strategies are not limited to English learners but are relevant wherever learners envision an aspirational linguistic identity---whether in acquiring global languages or in heritage language contexts where issues of identity and belonging are equally salient. 
By situating AI-mediated feedback within these motivational theories, we demonstrate how self-clone personalization can complement existing pedagogical practices in SLA and open new opportunities for identity-oriented language support. 
At the same time, our system is constrained by its reliance on LLMs, ASR, and voice synthesis, which are heavily dependent on data availability. 
For low-resource languages, these components may underperform or remain unavailable, limiting the reach of our approach and raising questions about equity in access to personalized AI learning technologies. 
Future research should therefore investigate how personalization interacts with linguistic distance, proficiency level, and cultural context, as well as how technical infrastructures can be advanced to support learners of under-resourced languages.

\subsection{Design Implications of Aspirational AI Self-Clones for HCI}
Beyond the context of language learning, AI Twin illustrates how AI self-clones can shape emotional engagement by aligning system output with learners' identities and aspirations. While prior works on AI self-clones have emphasized high-fidelity mimicry, such as reproducing an individual's voice, style, or conversational patterns ~\cite{leong2024dittos, huang2025mirror}, they rarely consider the role of aspirational selves. Drawing on the theoretical lens of the \emph{Ideal L2 Self}, our work shows how modeling a more fluent, confident version of the learner can have positive motivational effects. This suggests that the design space of AI self-clones extends beyond simple replication: rather than only mirroring the current self, self-clones can be designed to embody an aspirational or improved version of the user. Such aspirational self-clones could be valuable not only in language learning, but also in adjacent fields where motivation, self-perception, and affective support are critical---such as mental health interventions, professional skills training, or physical rehabilitation.

Building on these conceptual insights, our findings highlight opportunities for extending how aspirational self-clones might be designed in practice. While our prototype focused on voice, aspirational self-representation extends to traits such as personality and interactional style. Enabling users to specify which qualities to amplify can help ensure that the aspirational persona feels more aligned with their goals and identity.

However, aspiration cannot always be prioritized in isolation; it introduces tradeoffs with the system’s broader goals---in our case, supporting learning. From a learning perspective, feedback becomes a key site where these aims intersect. Participants differed in their preferences for implicit versus explicit correction, suggesting that no single feedback style serves both aspirational alignment and pedagogical clarity. This points to an opportunity for extending the design of AI Twin with adjustable or hybrid feedback mechanisms, allowing learners to tune how strongly the system foregrounds aspiration versus direct guidance.

Taken together, these considerations point toward broader implications for HCI. They underscore the importance of examining personalization not only as a matter of usability or performance optimization, but as a mechanism that shapes deeper psychological dynamics of self-perception, motivation, and identity. Designing systems that balance accurate self-representation with aspirational modeling opens new opportunities for affectively supportive, motivationally aligned human–AI interaction.

\subsection{Ethical Considerations}
Our approach demonstrates the motivational potential of voice-cloned, self-representational AI, but these opportunities must be weighed against substantive ethical risks~\cite{floridi2018ai4people, jobin2019global}. Voice cloning technologies inherently raise concerns related to consent, misuse, and data security~\cite{amezaga2022availability}. In our study, participants were fully informed about how their voice data would be collected and used, but broader deployment of such systems will require even more rigorous and transparent consent processes. Clear communication about data handling practices---including storage, processing, retention, and deletion---is essential, as is preparing users for risks that extend beyond the immediate learning context. Without such safeguards, voice models could be vulnerable to misuse, unauthorized replication, or unintended exposure~\cite{hutiri2024not}, underscoring the need for strong protection measures when adopting voice cloning in educational technologies.

Beyond risks inherent to voice cloning, self-representational AI systems raise additional concerns around how personal identity is represented and shaped~\cite{mcilroy2022mimetic}.
Recent work warns that AI-generated self-clones can threaten users' sense of authenticity~\cite{liu2025social} and may evoke negative reactions when they appear to exploit or displace one's identity, a phenomenon Lee et al.~\cite{lee2023speculating} describe as \emph{doppelganger-phobia}. They also highlight risks of \emph{identity fragmentation}, where multiple AI replicas complicate a user's cohesive self-perception.

These issues are relevant to our system, which generates an idealized ``more fluent'' version of the user shaped by design decisions (e.g., how fluent or standard the voice should sound) and by the model's generative tendencies.
Although our goal is to support learners' Ideal L2 Selves, they may perceive an agent's identity in unanticipated ways~\cite{xiao2007role}, influencing how they imagine their future selves should sound. Because the system blends users' vocal timbre with standardized English accent, the resulting voice does not necessarily reflect the authentic linguistic identity they would naturally develop. This raises questions about authenticity and whether these systems implicitly promote certain norms. These concerns highlight the need for careful design that supports motivation without constraining learners' autonomy or shaping their linguistic identities.

\section {Limitations and Future Work}
While our findings highlight the promise of AI Twin as a supportive and motivational tool for ESL learners, several limitations constrain the generalization and interpretation of our results.

First, our study involved only a single brief interaction across the three conditions. While this design allowed us to compare their immediate impacts on engagement, it limited our ability to evaluate actual learning. Although AI Twin fostered positive engagement, the short interaction made it difficult to determine whether any feedback was internalized or led to measurable learning. This challenge was compounded by the absence of an explicit learning target, which further limited the evaluation of learning outcomes.
Prior literature also points to plausible risks associated with AI-supported learning, such as reduced critical thinking when learners rely heavily on AI tools~\cite{zhai2024effects, georgiou2025chatgpt}. These considerations highlight the value of examining not only engagement but also how learners process and apply AI-mediated feedback.
Incorporating explicit learning objectives and repeated practice opportunities in future work would enable finer-grained evaluation of learning processes and outcomes, making it possible to test whether the motivational boost observed with AI Twin translates into improved learning.

Second, the single-session format offered only a snapshot of learners' immediate reactions.
Order analysis in Section~\ref{subsection_findings_twin} showed that encountering AI Twin first did not amplify its benefits, yet the intrinsic novelty of hearing one's cloned voice may still have contributed to the observed engagement regardless of condition order.
At the same time, participants attributed their experience to mechanisms beyond the novelty of the cloned voice---such as reduced anxiety about errors (``it was okay to make mistakes'', P17) and motivational responses consistent with Ideal L2 self-concepts (``I could speak like this too'', P13)---and none explicitly framed their reactions as novelty-driven.
Still, the single-session design limits our ability to distinguish transient novelty effects from more enduring engagement pathways.
Longitudinal studies would better clarify whether engagement benefits persist beyond initial exposure and whether they translate into sustained motivation, skill development, and language uptake over time.

Lastly, our participant pool consisted exclusively of Korean adult ESL learners, which limits the generalizability of our findings. As noted in our discussion, cultural and linguistic backgrounds influence how learners understand and make sense of self-representation technologies, shaping how they experience AI Twin. Evaluating the system with learners from a wider range of cultural contexts, age groups, and language backgrounds would improve ecological validity and help determine whether AI self-clones function similarly across diverse populations. Such expansion would also allow us to examine how different groups respond to various forms of feedback and personalization, and whether the motivational role of AI Twin holds in different sociocultural contexts.

\section{Conclusion}
This work introduced AI Twin, a personalized self-clone that reformulates learner utterances into more fluent versions and delivers them in the learner's own voice. 
Our within-subject study with 20 adult ESL learners demonstrated that this design fostered significantly higher emotional engagement than explicit correction and enhanced motivation by aligning feedback with learners' Ideal L2 Selves. 
Participants described AI Twin as more enjoyable, reassuring, and immersive, and many reported that hearing a more fluent version of themselves strengthened their confidence and desire to continue learning. 
These findings highlight the promise of self-representative AI in supporting affective and motivational dimensions of language learning that are often overlooked by systems focused solely on accuracy.

At the same time, our study surfaced nuances in how learners balance engagement with perceptions of learning, and in how they orient toward self-clone versus role-model representations. Rather than detracting from the approach, these variations highlight the opportunity to design flexible systems that accommodate diverse identity-related and motivational needs. Beyond ESL, AI Twin illustrates how personalization can extend into identity and affect, opening new possibilities for HCI to create inclusive, motivationally aligned technologies.
We seek to encourage broader exploration of self-representative AI, particularly within learning technologies, to design systems that are not only effective but also motivating and affectively supportive.

\bibliographystyle{ACM-Reference-Format}
\bibliography{main_camera_ready}

\newpage
\onecolumn
\appendix
\section{Prompts}
\label{appendix:prompts}
All prompts used in our study are presented in this section. The message generation prompt (Section~\ref{prompt_chat}) and the task completion tracking prompt (Section~\ref{prompt_task_completion}) were employed consistently across all experimental conditions. 
The AI Twin/Proxy prompt (Section~\ref{prompt_twin}) was shared between the \emph{AI Twin} and \emph{AI Proxy} conditions, 
while the Explicit Feedback prompt (Section~\ref{prompt_explicit_feedback}) was used exclusively in the \emph{Explicit Feedback} condition. 
\textcolor{blue}{Blue text} marks inputs that were programmatically supplied, and \textcolor{orange}{orange text} highlights outputs generated by the LLM.

\subsection{Chat}
\label{prompt_chat}
\hrulefill
\begin{Verbatim}[commandchars=\\\{\}]
You are a friendly and engaging conversation partner. 
The user is an ESL learner who wants to practice casual, spoken English. 
In this scenario, the user is in a specific everyday situation. 
Help them speak naturally and confidently, while gently encouraging them to express key ideas 
relevant to the context.

[CONTEXT]
\textcolor{blue}{\{\{mission_context\}\}}

[TASK]
- NEVER directly state the goals listed above. Gently guide the user toward them.
- NEVER rush through the goals; let the user explore their thoughts in full sentences.
- NEVER ask double-barrelled questions (ALWAYS ask one question at a time).
- NEVER ask yes/no questions.
- Make sure to keep the response short—up to two sentences.
- Make sure the interaction feels realistic, friendly, and enjoyable.

[DIALOGUE]
\textcolor{blue}{\{\{dialogue\}\}}
\end{Verbatim}
\hrulefill

\subsection{Task Completion Tracking}
\label{prompt_task_completion}
\hrulefill
\begin{Verbatim}[commandchars=\\\{\}]
Evaluate the following [DIALOGUE] between USER and ASSISTANT. 
Determine whether the USER has successfully completed each goal in the [MISSION].

[MISSION]
\textcolor{blue}{\{\{mission_context\}\}}

[OUTPUT FORMAT]
Respond with a valid JSON object containing the task completion status for each goal. 
The JSON should have the following structure:
\textcolor{orange}{ \{ }
    \textcolor{orange}{"task_results": [ }
    \textcolor{orange}{\{ "goal_number": 1, "completed": true \},}
    \textcolor{orange}{\{ "goal_number": 2, "completed": false \},}
    \textcolor{orange}{...}
  \textcolor{orange}{]}
\textcolor{orange}{\}}

[DIALOGUE]
\textcolor{blue}{\{\{dialogue\}\}}
\end{Verbatim}
\hrulefill

\subsection{AI Twin/Proxy}
\hrulefill
\label{prompt_twin}
\begin{Verbatim}[commandchars=\\\{\}]
You are the user's AI Twin/Proxy.
Your role is to support the user's English learning by rephrasing their utterances only when 
necessary to improve grammar or naturalness. You will be given recent dialogue history in
[DIALOGUE] and should rephrase only the user's most recent message following the instructions in
[TASK].

[TASK]
- Rephrase the user's sentence only if it contains grammar or usage mistakes.
- If the sentence is grammatically correct and sounds natural, repeat it as-is with no changes.
- When rephrasing, preserve the original meaning and tone of the user's message.
- Do not explain the correction. Only output the corrected version, nothing else.
- NEVER EVER answer user's question 
    - e.g., when user says "thank you", you are supposed to repeat "thank you", not "you're
            welcome".
- You are only allowed to speak in English.

[DIALOGUE]
\textcolor{blue}{\{\{dialogue\}\}}
\end{Verbatim}
\hrulefill

\subsection{Explicit Feedback}
\label{prompt_explicit_feedback}
\hrulefill
\begin{Verbatim}[commandchars=\\\{\}]
You are a helpful English teacher. The user is a Korean ESL learner engaging in an English
conversation. Follow the instructions in [TASK].

[TASK]
- This is feedback for a speaking session, so focus on how the user speaks.
- Give feedback on any grammatical, vocabulary, or expression errors in the user's English
  messages, focusing especially on natural, colloquial usage.
- Be concise.
- Use the recent dialogue history (provided) to understand the context and ensure accurate
  corrections.
- Ignore all spelling and capitalization errors. Treat them as if they were correct, and do not
  comment on them.
- Explain the corrections clearly in Korean, so the user can understand why the change is
  necessary.
- Make sure the feedback is clear and concise; avoid including the full correction.
- Use HTML formatting for emphasis: <strong>text</strong> for important corrections and <br> for
  line breaks.

[EXAMPLE]
- Sample input
It's my first time to visit here so can I ask you something about menu?
- Sample Feedback
문장을 두 부분으로 나누는 게 더 자연스러워요. 
<br><strong>"to visit"</strong> 대신에 <strong>"visiting"</strong>이 더 적절한 형태이고, 
<strong>"menu"</strong> 앞에 <strong>"the"</strong>를 붙이는 게 좋아요.

[DIALOGUE]
\textcolor{blue}{\{\{dialogue\}\}}
\end{Verbatim}
\hrulefill

\newpage
\section{Tasks}
\label{appendix:tasks}

For our conversational system implementation, we adopted a goal-oriented conversation structure to ensure methodological consistency across participants and to scaffold interactions in a supportive manner. 
Each session was framed as a role-play scenario in which the USER pursued a set of clearly defined communicative goals, thereby reducing hesitation and enabling controlled yet realistic conversational practice. 
The six tasks employed in the study are summarized in Table~\ref{table:tasks}. To discourage direct copying, task contexts and objectives were presented to participants in Korean. 
Each conversation was initiated by a standardized Initial Question, which established the scenario and provided a natural conversational opening.

\begin{table}[h]
\small
\centering
\caption{Conversational tasks used in the study, with context, goals, and standardized initial question.}
\label{table:tasks}
\renewcommand{\arraystretch}{1.2}
\resizebox{\linewidth}{!}{%
\begin{tabular}{p{0.34\linewidth} p{0.40\linewidth} p{0.26\linewidth}}
\hline
\textbf{Context} & \textbf{Goals} & \textbf{Initial Question} \\ \hline

\textbf{Task 1. Restaurant Reservation:} \newline Calling a Thai restaurant to plan a surprise birthday dinner. &
\begin{tabular}[t]{@{}l@{}}
-- Make a reservation for 6 people (Saturday, 7 pm).\\
-- Ask about reserving a private room.\\
-- Ask about bringing an outside birthday cake.\\
-- Inquire about parking availability.
\end{tabular} &
``Hello! Thank you for calling Siam Orchid. How can I help you today?'' \\
\hline

\textbf{Task 2. Dining Out with Friend} \newline Ordering with a vegetarian friend at a restaurant. &
\begin{tabular}[t]{@{}l@{}}
-- State friend is vegetarian and ask about options.\\
-- Inform waiter of nut allergy.\\
-- Ask for a drink recommendation.\\
-- Place final order.
\end{tabular} &
``Hi there! What can I get for you today?'' \\
\hline

\textbf{Task 3. Movie Planning} \newline Deciding on a movie with a friend. &
\begin{tabular}[t]{@{}l@{}}
-- Suggest a movie.\\
-- Explain interest or anticipated scenes.\\
-- Propose or agree on a time.\\
-- Decide when and where to meet.
\end{tabular} &
``Hey! Have you thought about which movie you want to watch?'' \\
\hline

\textbf{Task 4. Café Visit} \newline Ordering a drink and asking about amenities. &
\begin{tabular}[t]{@{}l@{}}
-- Ask about seasonal or new menu items.\\
-- Ask if plant-based milk options are available.\\
-- Request the Wi-Fi password.\\
-- Ask where power outlets are for phone charging.
\end{tabular} &
``Welcome! What kind of drink can I get started for you?'' \\
\hline

\textbf{Task 5. Hotel Check-In:} \newline Checking in at a hotel and confirming room details. &
\begin{tabular}[t]{@{}l@{}}
-- Confirm hotel reservation.\\
-- Ask whether the room has a nice view.\\
-- Ask if breakfast is included.\\
-- Ask about check-out time and procedures.
\end{tabular} &
``Good afternoon! Welcome to Maplewood Hotel. How may I help you today?'' \\
\hline

\textbf{Task 6. Furniture Shopping:} \newline Speaking with a salesperson at a furniture store. &
\begin{tabular}[t]{@{}l@{}}
-- Describe desired sofa style or design.\\
-- Ask if delivery is available by next week.\\
-- Ask about available color options.\\
-- Ask about sales or promotions.
\end{tabular} &
``Hi! Is there anything specific you're looking for today?'' \\
\hline

\end{tabular}}
\end{table}

\newpage
\section{Per-Question Post-hoc Analyses}
\label{appendix:per_question_post_hoc}
This section provides per-question pairwise post-hoc comparisons for all 10 questionnaire items that showed significant differences in the ANOVA (9 emotional engagement items and 1 behavioral engagement item). Table~\ref{tab:per_question_post_hoc} reports paired t-tests between the three feedback conditions, including the test statistic ($t$), corresponding p-values, and within-subjects effect sizes ($d_z$).

\newcolumntype{P}[1]{>{\centering\arraybackslash}p{#1}}
\renewcommand{\arraystretch}{1.2}
\begin{table}[h]
\centering
\caption{Pairwise post hoc comparisons between conditions for engagement (paired $t$-tests). Reported values are the test statistic ($t$), $p$-value, and within-subjects effect size ($d_z$).}
\begin{tabular}{P{1cm}|P{5.5cm}ccc}
\hline
\rowcolor{gray!30}
\multicolumn{1}{c}{\#} &
\multicolumn{1}{c}{\textbf{Comparison}} &
\multicolumn{1}{c}{\textbf{$t$}} &
\multicolumn{1}{c}{\textbf{$p$}} &
\multicolumn{1}{c}{\textbf{$d_z$}} \\ \hline
\rowcolor{gray!15}
\multicolumn{5}{l}{\textbf{Emotional Engagement}} \\ \hline
\multirow{3}{*}{1} & AI Proxy vs Explicit Feedback & 2.405 & \textbf{0.0265*} & 0.538 \\
                & AI Proxy vs AI Twin & -0.890 & 0.3847 & -0.199 \\
                & Explicit Feedback vs AI Twin & -2.854 & \textbf{0.0102*} & -0.638 \\ \hline
\multirow{3}{*}{2} & AI Proxy vs Explicit Feedback & 3.000 & \textbf{0.0074**} & 0.671 \\
                & AI Proxy vs AI Twin & -1.377 & 0.1845 & -0.308 \\
                & Explicit Feedback vs AI Twin & -3.039 & \textbf{0.0068**} & -0.679 \\ \hline
\multirow{3}{*}{3} & AI Proxy vs Explicit Feedback & 3.489 & \textbf{0.0025**} & 0.780 \\
                & AI Proxy vs AI Twin & -0.940 & 0.3590 & -0.210 \\
                & Explicit Feedback vs AI Twin & -3.199 & \textbf{0.0047**} & -0.715 \\ \hline
\multirow{3}{*}{5} & AI Proxy vs Explicit Feedback & 1.484 & 0.1543 & 0.332 \\
                & AI Proxy vs AI Twin & -1.917 & 0.0705 & -0.429 \\
                & Explicit Feedback vs AI Twin & -2.651 & \textbf{0.0158*} & -0.593 \\ \hline
\multirow{3}{*}{6} & AI Proxy vs Explicit Feedback & 4.802 & \textbf{0.0001***} & 1.074 \\
                & AI Proxy vs AI Twin & 1.045 & 0.3092 & 0.234 \\
                & Explicit Feedback vs AI Twin & -2.814 & \textbf{0.0111*} & -0.629 \\ \hline
\multirow{3}{*}{8} & AI Proxy vs Explicit Feedback & 3.269 & \textbf{0.0040**} & 0.731 \\
                & AI Proxy vs AI Twin & 0.000 & 1.0000 & 0.000 \\
                & Explicit Feedback vs AI Twin & -2.565 & \textbf{0.0190*} & -0.573 \\ \hline
\multirow{3}{*}{10} & AI Proxy vs Explicit Feedback & 2.998 & \textbf{0.0074**} & 0.670 \\
                & AI Proxy vs AI Twin & 1.437 & 0.1670 & 0.321 \\
                & Explicit Feedback vs AI Twin & -2.032 & 0.0563 & -0.454 \\ \hline
\multirow{3}{*}{11} & AI Proxy vs Explicit Feedback & 4.076 & \textbf{0.0006***} & 0.911 \\
                & AI Proxy vs AI Twin & 0.490 & 0.6295 & 0.110 \\
                & Explicit Feedback vs AI Twin & -3.884 & \textbf{0.0010**} & -0.869 \\ \hline
\multirow{3}{*}{12} & AI Proxy vs Explicit Feedback & 2.431 & \textbf{0.0251*} & 0.544 \\
                & AI Proxy vs AI Twin & 0.181 & 0.8582 & 0.041 \\
                & Explicit Feedback vs AI Twin & -2.761 & \textbf{0.0124*} & -0.617 \\ \hline
\rowcolor{gray!15}
\multicolumn{5}{l}{\textbf{Behavioral Engagement}} \\ \hline
\multirow{3}{*}{23} & AI Proxy vs Explicit Feedback & -2.027 & 0.0569 & -0.453 \\
                & AI Proxy vs AI Twin & 1.129 & 0.2731 & 0.252 \\
                & Explicit Feedback vs AI Twin & 2.319 & \textbf{0.0317*} & 0.519 \\ \hline
\end{tabular}
\label{tab:per_question_post_hoc}
\end{table}

\end{document}